\documentclass[acmsmall]{acmart}
\usepackage{color}
\usepackage{multirow}
\usepackage{xspace}
\usepackage{algorithm}
\usepackage{algpseudocode}
\usepackage[normalem]{ulem}
\useunder{\uline}{\ul}{}
\newcommand{\modelname}{FELLAS\xspace} 
\newcommand{\modelnamenospace}{FELLAS}

\AtBeginDocument{%
  }

\setcopyright{acmlicensed}
\copyrightyear{2018}
\acmYear{2018}
\acmDOI{XXXXXXX.XXXXXXX}

\acmJournal{JACM}
\acmVolume{37}
\acmNumber{4}
\acmArticle{111}
\acmMonth{8}

\begin{document}

\title{\modelname: Enhancing Federated Sequential Recommendation with LLM as External Services}

\author{Wei Yuan}
\email{w.yuan@uq.edu.au}
\affiliation{%
  \institution{The University of Queensland}
  \city{Brisbane}
  \state{QLD}
  \country{Australia}
}

\author{Chaoqun Yang}
\email{chaoqun.yang@griffith.edu.au}
\affiliation{%
  \institution{Griffith University}
  \city{Gold Coast}
  \state{QLD}
  \country{Australia}
}

\author{Guanhua Ye}
\email{g.ye@bupt.edu.cn}
\affiliation{%
 \institution{Beijing University of Posts and Telecommunications}
 \city{Beijing}
 \state{Beijing}
 \country{China}}

 \author{Tong Chen}
\email{tong.chen@uq.edu.au}
\affiliation{%
  \institution{The University of Queensland}
  \city{Brisbane}
  \state{QLD}
  \country{Australia}
}

 \author{Quoc Viet Hung Nguyen}
\email{henry.nguyen@griffith.edu.au}
\affiliation{%
  \institution{Griffith University}
  \city{Gold Coast}
  \state{QLD}
  \country{Australia}
}

\author{Hongzhi Yin}\authornote{Corresponding author.}
\email{h.yin1@uq.edu.au}
\affiliation{%
  \institution{The University of Queensland}
  \city{Brisbane}
  \state{QLD}
  \country{Australia}
}

\renewcommand{\shortauthors}{Yuan et al.}

\begin{abstract}
  Sequential recommendation has been widely studied in the recommendation domain since it can capture users' temporal preferences and provide more accurate and timely recommendations.
  To address user privacy concerns, the combination of federated learning and sequential recommender systems (FedSeqRec) has gained growing attention.
  Unfortunately, the performance of FedSeqRec is still unsatisfactory because the models used in FedSeqRec have to be lightweight to accommodate communication bandwidth and clients' on-device computational resource constraints.
  Recently, large language models (LLMs) have exhibited strong transferable and generalized language understanding abilities and therefore, in the NLP area, many downstream tasks now utilize LLMs as a service to achieve superior performance without constructing complex models.
  Inspired by this successful practice, we propose a generic FedSeqRec framework, \modelname, which aims to enhance FedSeqRec by utilizing LLMs as an external service.
  
  Specifically, \modelname employs an LLM server to provide both item-level and sequence-level representation assistance. 
  The item-level representation service is queried by the central server to enrich the original ID-based item embedding with textual information, while the sequence-level representation service is accessed by each client. 
  However, invoking the sequence-level representation service requires clients to send sequences to the external LLM server. 
  To safeguard privacy, we implement $d_{\mathcal{X}}$-privacy satisfied sequence perturbation, which protects clients' sensitive data with guarantees.
  Additionally, a contrastive learning-based method is designed to transfer knowledge from the noisy sequence representation to clients' sequential recommendation models.
  Furthermore, to empirically validate the privacy protection capability of \modelname, we propose two interacted item inference attacks, considering the threats posed by the LLM server and the central server acting as curious-but-honest adversaries in cooperation.
  Extensive experiments conducted on three datasets with two widely used sequential recommendation models demonstrate the effectiveness and privacy-preserving capability of \modelname. 
\end{abstract}



\received{20 February 2007}
\received[revised]{12 March 2009}
\received[accepted]{5 June 2009}

\maketitle

\section{Introduction}
The demand for recommender systems has increased in online services since they can effectively alleviate information overload~\cite{chen2023deep}.
As a new branch of recommender systems, sequential recommender systems capture users' dynamic preferences and provide potential recommendations from massive candidates by modeling users' historical interactions as temporally ordered sequences. 
Therefore, they are widely used on online platforms (e.g., e-commerce~\cite{wang2019sequential,guo2021hierarchical}) to alleviate information overload. 
Traditionally, developing an effective sequential recommender system requires collecting vast amounts of users' private interaction data on a central server, posing significant risks of data leakage and privacy concerns~\cite{wang2022trustworthy}. 
With the recent release of privacy protection regulations in many regions (e.g., GDPR~\cite{voigt2017eu} in the EU and CCPA~\cite{harding2019understanding} in the USA), this centralized, privacy-unaware training paradigm has become increasingly challenging to apply without violating regulations~\cite{yuan2023interaction}.

To address privacy concerns, researchers propose adopting federated learning~\cite{zhang2021survey}, a privacy-preserving learning scheme, to train a sequential recommender system known as the federated sequential recommender system (FedSeqRec)~\cite{lin2022generic}. 
In FedSeqRec, clients/users\footnote{In this paper, we focus on user-level federated sequential recommendation, where the terms "user" and "client" are used interchangeably as one client is one user.} can collaboratively train a sequential recommendation model without exposing their private data. Instead, users only need to transmit model parameters under the coordination of a central server. As a result, users' data privacy is protected.

Recent research on leveraging large language models (LLMs) to enhance sequential recommendation performance has achieved initial success~\cite{zhao2024recommender} and quickly become a popular topic. However, most of these studies are conducted in a centralized paradigm and do not address user data privacy. To bridge this gap, some researchers~\cite{zhao2024llm,zhang2024federated} have begun exploring the integration of FedSeqRec with LLMs. These works primarily treat LLMs as the backbone of sequential recommender systems and fine-tune them within the FedSeqRec framework.
While these approaches use ``efficient fine-tuning'' methods like LoRA~\cite{hu2021lora}, they remain impractical due to several limitations. First, despite efficient fine-tuning, communication costs are still significantly increased. Furthermore, the computational resources available to clients (e.g., on-device GPUs and battery power) are insufficient to support LLM training. Although Zhao et al.~\cite{zhao2024llm} address this issue by using split training, the communication overhead becomes much worse since clients must transfer intermediate results for each forward and backward pass.
In addition, relying on LLMs as the core model leads to considerable inference delays, a problem that worsens when these models are deployed on users' low-powered devices.

In the natural language processing (NLP) community, large language models (LLMs)~\cite{brown2020language,min2023recent} have exhibited exceptional and generalized capabilities in natural language understanding.
Consequently, numerous academic studies~\cite{sun2022black} and industrial applications (e.g., GPT series APIs in OpenAI\footnote{\url{https://openai.com/index/openai-api/}}) have explored leveraging LLMs to provide third-party services, such as embeddings, to assist customers in various NLP tasks.
By using embeddings provided by advanced LLMs, customers can achieve good performance in downstream NLP tasks even with simple models~\cite{peng2023you}.

\begin{figure}[!t]
  \centering
  \includegraphics[width=0.9\linewidth]{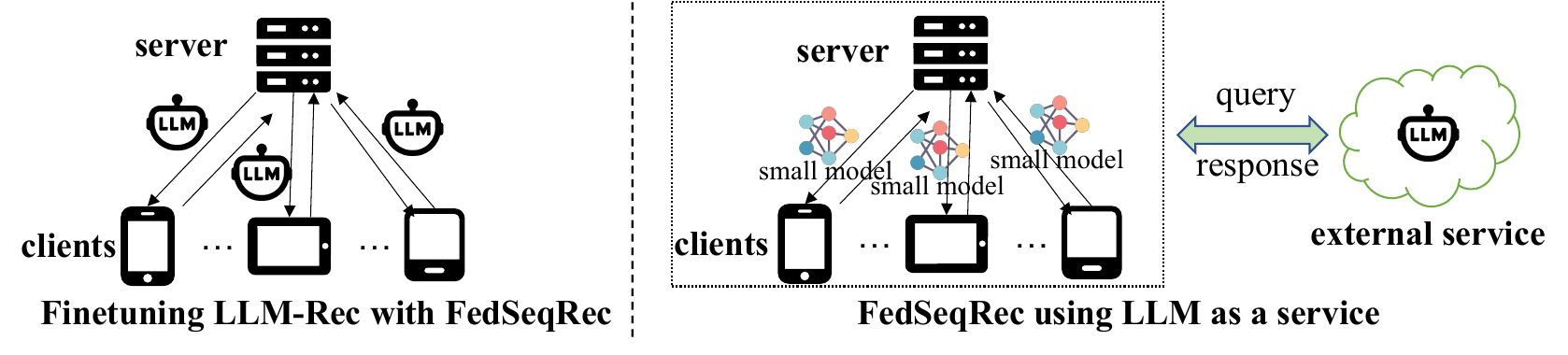}\caption{The comparison of finetuning LLM in FedSeqRec and using LLM as a service in FedSeqRec.}
  \label{fig_fsq_vs_service}
  \end{figure}

In light of the successful practice of using LLMs as a service in NLP, we propose a novel and generic framework, \modelname (\underline{F}ederated S\underline{e}quential Recommendation Framework with \underline{LL}M \underline{a}s External \underline{S}ervice), to explore the utilization of a pretrained LLM as an external service to remedy the poor representation ability of the sequential recommendation model in FedSeqRec.
Figure~\ref{fig_fsq_vs_service} highlights the differences between fine-tuning an LLM as a recommender and using an LLM as an external service. Compared to recent works that directly employ LLMs as recommenders in federated recommendation~\cite{zhao2024llm}, our \modelname offers several key advantages.
First, \modelname does not introduce significant communication overhead, as the basic training flow of FedSeqRec remains unchanged. Additionally, computational costs are not substantially increased, since clients continue training with traditional sequential recommender systems, which are much more lightweight than LLM-based models. Finally, \modelname maintains the inference speed of traditional sequential recommendation models while leveraging the powerful representational abilities of LLMs.
As a result, \modelname offers a more practical approach.

\begin{figure}[!t]
  \centering
  \includegraphics[width=0.9\linewidth]{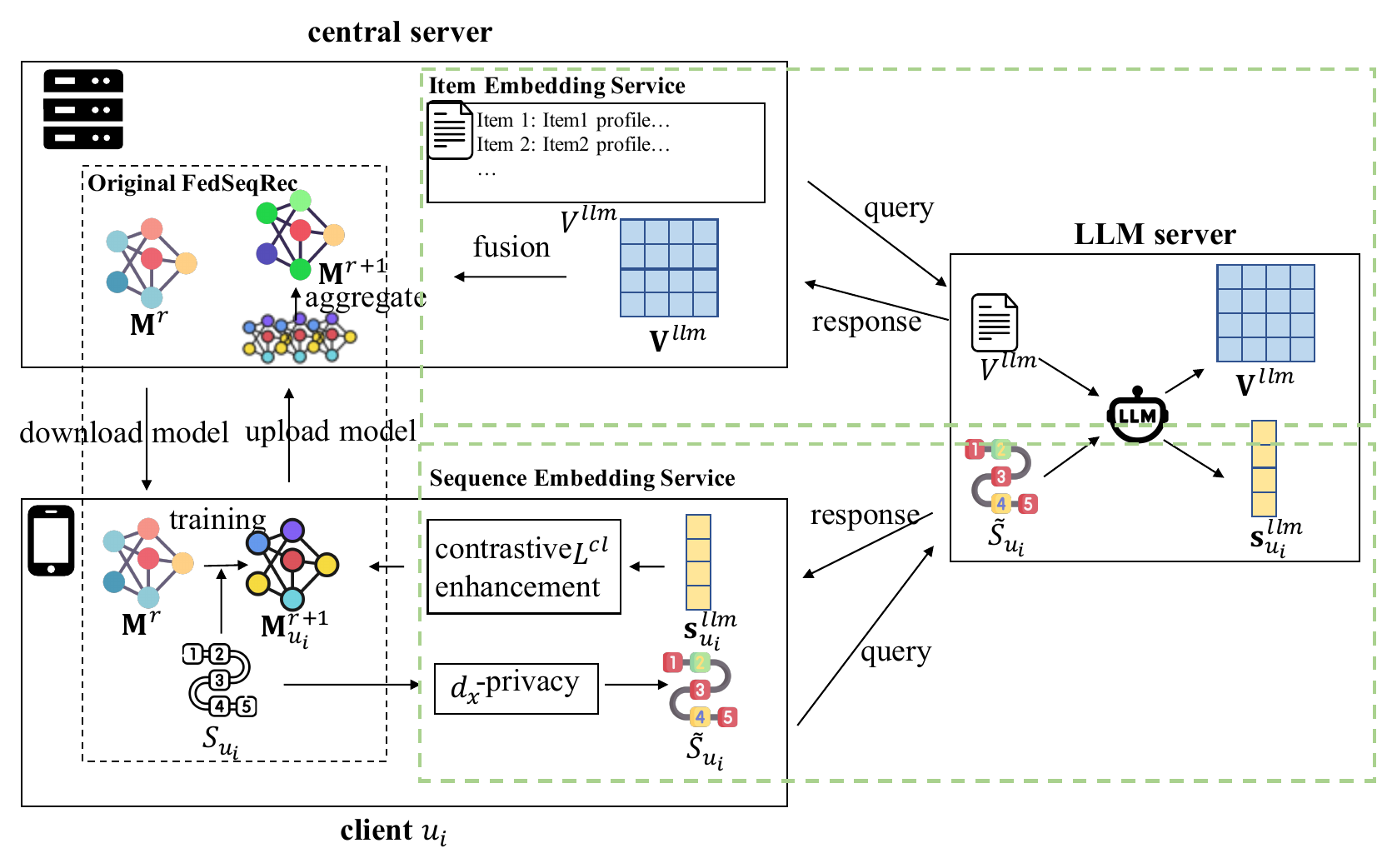}\caption{The overview of \modelname. The black dashed line box highlights the original FedSeqRec while the green dashed line boxes are LLM services incorporated in \modelname.}
  \label{fig_arch}
  \end{figure}

Figure~\ref{fig_arch} presents the details of our \modelname architecture. 
In \modelname, there is an LLM server that provides two types of representation services. 
The first is the item-level representation service, which is used by the central server. 
Specifically, the central server sends the items' textual information to the LLM server and obtains textual item embeddings. 
These textual embeddings are then fused with the original ID-based item embeddings to enhance recommendation performance.

The second representation service is to provide the sequence-level embedding for the clients.
However, it is non-trivial to adopt this service in FedSeqRec, since clients need to upload interaction sequences to the LLM server for encoding, which would break the basic FedSeqRec learning protocol and finally compromise user privacy.
To address this challenge, we employ a $d_{\mathcal{X}}$-privacy~\cite{chatzikokolakis2013broadening} based perturbation method to construct noisy but meaningful sequences based on their raw data at each client.
Then, clients send the noisy sequences to the LLM server to obtain a noisy version of sequence representations. 
Subsequently, a contrastive learning method is developed to transfer sequential knowledge contained in the noisy sequence representation to enhance clients' local training.

Furthermore, to validate the privacy-preserving ability of FedSeqRec after incorporating LLM as a service, we propose two interacted item inference attacks each considering a different threat scenario: 
(1) a similarity-based inference attack (SIA) that assumes the external LLM server is a curious-but-honest adversary, and (2) a similarity-based inference attack with update information (SIAUI) that assumes both the LLM server and the central server are curious-but-honest adversaries engaged in malicious cooperation. 
It is worth noting that there is another potential threat scenario in \modelname: the central server acting as a curious-but-honest attacker. However, this scenario is not unique to \modelname and many studies~\cite{lin2022generic,yuan2023interaction} have explored defenses to address this threat. \modelname can seamlessly integrate these solutions. Therefore, it is unnecessary to discuss this scenario in this paper.
As a result, by launching these two attacks, we can empirically validate and analyze the privacy protection capabilities of our \modelname.

To demonstrate the effectiveness of \modelname, we conducted extensive experiments on three widely used recommendation datasets and two sequential recommendation models. The experimental results indicate that \modelname can significantly improve the performance of FedSeqRec while maintaining its privacy-preserving capabilities.

To sum up, the major contributions of this paper are as follows:
\begin{itemize}
  \item To the best of our knowledge, we are the first to explore integrating LLM as a service to improve federated sequential recommendation performance.
  \item We propose a novel and generic framework, \modelname, which incorporates two representation services provided by LLM: item-level textual embedding and sequence-level embedding. When using the sequence-level embedding service, a perturbation method is employed to provide privacy guarantees, and a contrastive learning method is designed to effectively utilize the noisy sequence representations.
  \item We propose two interacted item inference attacks for two potential threat scenarios in \modelnamenospace: the external LLM server acting as a curious-but-honest adversary, and the LLM server cooperating with the central server to launch the attacks. These attacks validate the empirical privacy-preserving performance of \modelname.
  \item We conduct extensive experiments on three datasets collected from Amazon (Beauty, Office Products, and Patio Lawn and Garden)~\cite{ni2019justifying} with several FedSeqRecs, demonstrating the effectiveness of our proposed framework.
\end{itemize}

The remainder of this paper is organized as follows. 
The related works of sequential recommender systems, federated recommendations, contrastive learning in recommender systems, and LLMs as services are presented in Section~\ref{sec_related_work}.
Section~\ref{sec_preliminary} provides the preliminaries related to our research, including the problem definition of federated sequential recommender systems and basic sequential recommender systems.
Then, in Section~\ref{sec_methodology}, we present the technical details of our proposed \modelname, including item-level LLM service and privacy-preserving sequence-level LLM service, followed by two inference attack methods.
The experimental results with comprehensive analysis are exhibited in Section~\ref{sec_experiments}.
Finally, Section~\ref{sec_conclusion} gives a brief conclusion of this paper and provides some future research directions.

  \section{Related Work}\label{sec_related_work}
  In this section, we present the literature reviews for four related topics: sequential recommendation~\cite{wang2019sequential}, federated recommendation~\cite{yang2020federated,sun2022survey,yin2024device}, contrastive learning in recommendation~\cite{yu2023self}, and LLM as service~\cite{gan2023model}. 
  
  \subsection{Sequential Recommender System}
  In many online applications, temporal information in users' historical interactions is both valuable and crucial for modeling dynamic preferences and generating accurate recommendations~\cite{wang2019sequential}. 
  To leverage this temporal information, Rendle et al.~\cite{rendle2010factorizing} introduced one of the first sequential recommender systems, using a first-order Markov chain to model changing user preferences over time. With the advancement of deep learning, neural networks have since been widely adopted for sequential recommendation~\cite{fang2020deep}. 
  For example, Jannach et al.~\cite{jannach2017recurrent} utilized gated recurrent units (GRUs) as the backbone to encode user interaction sequences. However, GRUs suffer from gradient explosion or vanishing problems and can only capture item information in a strictly sequential manner. To address this, Kang et al.~\cite{kang2018self} proposed using self-attention mechanisms to learn user patterns, as these networks have a stronger ability to represent complex relationships.

  More recently, the success of large language models (LLMs) in natural language processing (NLP)~\cite{min2023recent} has inspired researchers to explore their potential for a sequential recommendation. Broadly, these efforts fall into two categories based on whether LLMs are used as the inference model. For instance, P5~\cite{geng2022recommendation} directly converts item indexes into text and feeds them with textual prompts into LLMs for fine-tuning. Similarly, Li et al.~\cite{li2023text} proposed Recformer by redesigning Longformer~\cite{beltagy2020longformer} and pretraining it on multiple recommendation datasets. CLLM4Rec~\cite{zhu2024collaborative} bridges the gap between NLP and recommendation by extending the LLM's vocabulary and employing a soft+hard prompting strategy. However, using LLMs as inference models for recommendation faces significant challenges, including high computational costs and severe inference delays, rendering these methods impractical.

  The second research line, ``LLMs as services'', typically uses LLMs to provide additional knowledge for traditional sequential recommendation models. For example, Ren et al.~\cite{ren2024representation} integrated LLMs into representation learning to capture intricate semantic aspects of user behaviors and preferences. Similarly, Wei et al.~\cite{wei2024llmrec} utilized LLMs to augment recommendation data. However, these methods do not address privacy concerns.

  \subsection{Federated Recommendation}  
  The combination of federated learning and recommender systems, known as federated recommender systems (FedRecs), has become a prominent topic in the recommendation research community due to its potential to address growing public concerns regarding data privacy in recommender systems~\cite{jeckmans2013privacy}.
  Ammand et al.~\cite{ammad2019federated} were the first to apply federated learning to collaborative filtering. 
  Since then, numerous studies have emerged to improve recommendation accuracy~\cite{lin2020fedrec,hung2017computing,liang2021fedrec++}, training efficiency~\cite{zhang2023lightfr}, and system trustworthiness~\cite{zhang2023comprehensive,yuan2023interaction}.
  Some work transplant FedRecs to specific recommendations, such as news recommendation~\cite{yi2021efficient}, social recommendation~\cite{liu2022federated}, POI recommendation~\cite{long2023decentralized}, and so on~\cite{huang2023incentive}.
  Additionally, some studies have explored incorporating advanced neural networks to enhance FedRecs, such as federated graph-based recommender systems~\cite{wu2021fedgnn} and federated sequential recommender systems~\cite{lin2022generic}.
  
  Recently, the development of large language models~\cite{li2021survey} has inspired researchers to harness their power in FedRecs.
  For example, Zhao et al.~\cite{zhao2024llm} straightforwardly utilized the current FedRec framework to train LLM-based recommender systems.
  Unfortunately, the vast number of parameters in LLMs compared to traditional recommender systems results in prohibitive communication costs and impractical computational resource requirements for clients.
  Zhang et al.~\cite{zhang2024federated} proposed FedPA, which uses pretrained LLMs to provide textual item embeddings for collaborative filtering-based recommendation tasks.
  In contrast to FedPA, our \modelname has at least two major differences, making them uncomparable:
  (1) \modelname not only uses LLMs to provide textual item embeddings but also explores using LLMs to encode user interaction sequences. This exploration is non-trivial as FedSeqRec does not allow users to expose their original data. Thus, \modelname implements a privacy-preserving sequence-sharing method and a contrastive learning method to learn from noisy sequence representations.
  (2) The tasks addressed by the two approaches are different: FedPA focuses on federated collaborative filtering, while \modelname is designed for a federated sequential recommendation.
  
  As a result, how to improve FedSeqRec with LLMs is still under-explored.
  In this paper, we innovatively utilize an LLM system as an external service and investigate how to leverage this external service to enhance traditional federated sequential recommender systems. This design can improve FedSeqRec performance while avoiding excessive communication and computational burdens.
  
  \subsection{Contrastive Learning in Recommendation}
  Contrastive learning has gained significant traction in recommender systems~\cite{yu2023self,zheng2023automl} by providing self-supervised signals.
  The core idea behind contrastive learning is to reduce the distance between positive instances while pushing negative instances farther apart in the representation space~\cite{tian2020makes}.
  For instance, S$^{3}$-Rec~\cite{zhou2020s3} introduces four auxiliary self-supervised objectives to capture correlations among attributes, items, subsequences, and sequences.
  CL4SRec~\cite{xie2022contrastive} applies contrastive learning in sequential recommendation.
  It executes item cropping, masking, and reordering on sequences to create different views for contrastive learning.

  Beyond augmenting data, contrastive learning has been applied to enhance model performance by multiple information view fusion.
  Wei et al.~\cite{wei2023multi} leveraged contrastive learning to fuse multi-modal information.
  KACL\cite{wang2023knowledge} performed contrastive learning between knowledge graph views and user-item interaction graphs to mitigate interaction noise.

  In this paper, we treat the sequence-level embedding from the LLM server as a contrastive view and design a contrastive learning method to fuse information into local sequential models.

  \subsection{Large Language Model as Service}  
  Large language models (LLMs) consist of numerous layers of neural networks, such as Transformers~\cite{vaswani2017attention}, with parameters ranging from millions to trillions. These models are trained on vast amounts of text through self-supervised or semi-supervised methods~\cite{zhao2023survey}. Generally, LLMs fall into two categories: discriminative and generative. BERT~\cite{devlin2019bert}, a well-known discriminative LLM, utilizes a bidirectional Transformer architecture and introduces masked language modeling for pretraining. Longformer~\cite{beltagy2020longformer} builds on this by incorporating an efficient, scalable self-attention mechanism that allows it to handle much longer input sequences than standard Transformer models like BERT.
  On the generative side, LLMs have seen remarkable success. GPT~\cite{radford2018improving} pretrains by predicting the next word in a sentence, a technique that powers many modern generative models. Notably, LLaMA2~\cite{touvron2023llama} and LLaMA3~\cite{dubey2024llama} are prominent open-source generative LLMs, with models ranging from 7 billion to 70 billion parameters.

  Advanced large language models~\cite{min2023recent} have demonstrated exceptional and broad language understanding capabilities.
  Consequently, the owners of LLMs have begun offering general-purpose APIs to provide basic services, such as embeddings and text generation~\cite{brown2020language,wang2021ernie,nguyen2017argument,gao2021making,zheng2023multi,meng2022generating}.
  This innovative and promising business model has quickly become common practice, benefiting both service providers and customers~\cite{gan2023model,peng2023you}. 
  Specifically, LLM service providers can generate profits to cover the costs of training LLMs, while customers can efficiently deploy powerful LLMs without the need for advanced AI infrastructure.

  In this work, we assume the existence of a pretrained LLM as an external service and explore how to utilize this system to improve traditional sequential recommender systems in a privacy-preserving manner.
  Specifically, we conduct experiments with the most representative LLMs including both discriminative LLMs (e.g., BERT and Longformer) and advanced generative LLMs (Llama2 and Llama3).

  \begin{table}[]
    \centering
    \caption{List of important notations.}\label{tb_notation}
    \begin{tabular}{l|l}
    \hline
     $\mathcal{S}_{u_{i}}$ & the local dataset (interacted item sequence) for user $u_{i}$.  \\
     $\widetilde{\mathcal{S}}_{u_{i}}$ & user $u_{i}$'s perturbed sequence.  \\
     $T_{u_{i}}$ & the length of user $u_{i}$'s interaction sequence.  \\
     \hline
     $\mathcal{U}$ & all users in the federated recommender system.  \\
     $\mathcal{U}^{r}$ & selected training users in $r$ round.  \\
     $\mathcal{V}$ & all items in the federated recommender system. \\
     $\mathcal{V}^{tx}$ & all items' textual information. \\
     $\mathcal{V}^{trained}_{u_{i}}$ & user $u_{i}$'s trained items. \\
     \hline
     $u_{i}$ & user $u_{i}$ is a user/client in federated recommendation system.   \\ 
     $v_{t}^{u_{i}}$ & the item that user $u_{i}$ interacted at time step $t$.   \\ 
     $v_{j}^{tx}$ & item $v_{j}$'s textual information.   \\ 
     $r_{v_{j},t}^{u_{i}}$ & the predicted score of user $u_{i}$ for item $v_{j}$ at $t$'th interaction.   \\ 
     $\mathbf{s}_{u_{i}}^{llm}$ & the sequence representation for $u_{i}$ generated by LLM server.   \\ 
     \hline
     $\mathbf{M}_{u_i}^{r}$ & user $u_{i}$'s model parameters in round $r$.   \\ 
     $\mathbf{M}^{r}$ & global model parameters in round $r$.   \\ 
     $\mathbf{V}$ & item embeddings.   \\ 
     $\mathbf{V}^{llm}$ & item embeddings generated by LLM server.   \\ 
     $\mathcal{F}^{llm}$ & LLM-based item encoder.   \\ 
     \hline
     $\alpha$ & the factor controls the strengths of contrastive learning.   \\
     $\epsilon$ & privacy parameter in $d_{\mathcal{X}}-$ privacy.   \\ 
     \hline
    \end{tabular}
    \end{table}

\section{Priliminaries}\label{sec_preliminary}
In this section, we present the basic background knowledge of federated sequential recommender systems and the base sequential recommendation models used in our work.
The notations used in this paper are under the following rules: the bold lowercases (e.g., $\mathbf{a}$) are vectors, the bold uppercases (e.g., $\mathbf{A}$) represent matrices or models, and the squiggle uppercase (e.g., $\mathcal{A}$) denotes sets or functions.
The meaning of the most important notations are listed in Table~\ref{tb_notation}.

\subsection{Federated Sequential Recommendation}\label{sec_base_fedseqrec}
Let $\mathcal{U} = \{u_{1}, u_{2}, \dots, u_{\left|\mathcal{U}\right|}\}$ and $\mathcal{V} = \{v_{1}, v_{2}, \dots, v_{\left|\mathcal{V}\right|}\}$ denote the sets of users and items in a sequential recommender system.
$\left|\mathcal{U}\right|$ and $\left|\mathcal{V}\right|$ represent the number of users and items, respectively. 
Each user $u_{i}$ has a chronological item interaction sequence $\mathcal{S}_{u_{i}} = [v_{1}^{u_{i}}, v_{2}^{u_{i}}, \dots, v_{t}^{u_{i}}, \dots, v_{T_{u_{i}}}^{u_{i}}]$, where $v_{t}^{u_{i}}$ is the item that user $u_{i}$ interacted with at time step $t$ and $T_{u_{i}}$ is the length of the sequence.
In federated sequential recommendation, to protect users' privacy, the raw interaction sequence $\mathcal{S}_{u_{i}}$ is stored on user $u_{i}$'s local devices.
Therefore, the goal of FedSeqRec is to learn a sequential recommendation model given the distributed user history logs, which can predict the item that $u_{i}$ will interact with at time step $T_{u_{i}} + 1$.

In FedSeqRec, a central server coordinates the clients to achieve the above goal.
Initially, the central server initializes a sequential model $\mathbf{M}^{0}$ and then deploys it on clients' devices.
Subsequently, clients and the central server collaboratively execute the following steps until model convergence.
At the start of round $r$, the central server selects a group of users $\mathcal{U}^{r}$ to participate in the training process and distributes recommendation model $\mathbf{M}^{r}$ to them.
Then, each user $u_{i} \in \mathcal{U}^{r}$ trains $\mathbf{M}^{r}$ with their interaction log $\mathcal{S}_{u_{i}}$ using a specific objective function $\mathcal{L}^{rec}$, such as binary cross entropy loss:
\begin{equation}\label{eq_ori_recloss}
  \mathcal{L}^{rec} = -\sum\limits_{v_{t}\in\mathcal{S}_{u_{i}}} \left[  log(\sigma (r_{v_{t},t})) + \sum\limits_{v_{j}\notin \mathcal{S}_{u_{i}}} log(1 - \sigma (r_{v_{j},t})) \right]
\end{equation}
where $v_{t}$ is the concise version of $v^{u_{i}}_{t}$,  $0\le r_{v,t}\le1$ is the prediction score of item $v$ at time step $t$ based on the sequence $\mathcal{S}_{u_{i}}$, indicating the preference of $u_{i}$ to $v$.       
After local training, user $u_{i}$ uploads the updated model $\mathbf{M}_{u_{i}}^{r+1}$ back to the central server.
The central server aggregates all received models to obtain the latest model $\mathbf{M}^{r+1}$.
Algorithm~\ref{alg_traditional} summarizes the vanilla federated sequential recommendation procedures in pseudo-code.

\begin{algorithm}[!ht]
  \renewcommand{\algorithmicrequire}{\textbf{Input:}}
  \renewcommand{\algorithmicensure}{\textbf{Output:}}
  \caption{The pseudo-code for vanilla federated sequential recommendation.} \label{alg_traditional}
  \begin{algorithmic}[1]
    \Require global round $R$; learning rate $lr$, \dots
    \Ensure  well-trained sequential model $\mathbf{M}^{R}$
    \State server initializes model $\mathbf{M}^{0}$
    \For {each round r =0, ..., $R-1$}
      \State sample a fraction of clients $\mathcal{U}^{r}$ from $\mathcal{U}$
        \For{$u_{i}\in \mathcal{U}^{r}$ \textbf{in parallel}} 
        \State // execute on client sides
        \State $\mathbf{M}_{u_{i}}^{r+1}\leftarrow$\Call{ClientTrain}{$u_{i}$, $\mathbf{M}^{r}$}
        \EndFor
      \State // execute on central server
      \State $\mathbf{M}^{r+1}\leftarrow$ aggregate received client model parameters $\{\mathbf{M}_{u_{i}}^{r+1}\}_{u_{i}\in \mathcal{U}^{r}}$
      \EndFor
    \Function{ClientTrain} {$u_{i}$, $\mathbf{M}^{r}$}
    \State $\mathbf{M}_{u_i}^{r+1}\leftarrow$ update local model with recommendation objective $\mathcal{L}^{rec}$
    \State \Return $\mathbf{M}_{u_{i}}^{r+1}$
    \EndFunction
    \end{algorithmic}
\end{algorithm}

\subsection{Base Sequential Recommender System}
The above FedSeqRec framework can be compatible with most current sequential recommendation models with reasonable sizes. 
In this paper, to demonstrate the generalization of our proposed \modelname, we select the two most popular sequential recommendation models, GRU4Rec~\cite{jannach2017recurrent} and SASRec~\cite{kang2018self} as the experimental base models.

\emph{GRU4Rec.} It leverages GRU~\cite{chung2014empirical} to capture the sequential information from users' interacted item logs.
Specifically, it takes a user interaction sequence $\mathcal{S} = [v_{1}, v_{2}, \dots, v_{t}]$ as input, and then, converts them into a sequence of vectors $[\mathbf{v}_{1}, \mathbf{v}_{2}, \dots, \mathbf{v}_{t}]$ using an item embedding table $\mathbf{V}$.
After that, a stack of GRU networks is applied in the embedding sequence and calculates a sequence of latent vectors $[\mathbf{h}_{1}, \mathbf{h}_{2}, \dots, \mathbf{h}_{t}]$.
Based on the latest latent vector $\mathbf{h}_{t}$, a feed-forward neural network is used to predict the user's preference scores at step $t+1$.
\begin{equation}
  \mathbf{r}_{*,t+1} = FFN(\mathbf{h}_{t})
\end{equation}

\emph{SASRec.} It uses self-attention to capture the relationships between items in a user's interaction sequence. This design allows the model to focus on the most relevant previous interactions when making predictions, irrespective of their distance in the sequence.
Since self-attention layers struggle with modeling the temporal order, SASRec employs positional embedding to retain the temporal information of the sequence.
Thus, for an input interaction sequence $\mathcal{S} = [v_{1}, v_{2}, \dots, v_{t}]$, SASRec converts it to vectors using both item embedding table and positional embedding table $[\mathbf{v}_{1} + \mathbf{p}_{1}, \mathbf{v}_{2} + \mathbf{p}_{2}, \dots, \mathbf{v}_{t} + \mathbf{p}_{t}]$.
Subsequently, several self-attention blocks with residual connections are stacked to compute the latent vectors $[\mathbf{h}_{1}, \mathbf{h}_{2}, \dots, \mathbf{h}_{t}]$.
Finally, the latest latent vector is used to calculate the user's preference scores on each item with the item embedding table:
 \begin{equation}
  \mathbf{r}_{*,t+1} = \mathbf{h}_{t}^\top \mathbf{V}
\end{equation}

\section{Methodology}\label{sec_methodology}
In this section, we first introduce the details of our proposed framework, \modelname. Following this, we propose two inference attacks to evaluate \modelname's empirical privacy protection capabilities, given that users in \modelname will upload a noisy version of their data to a third-party LLM server.

\subsection{Overview of \modelname}
To address the poor representation ability of sequential recommendation models in FedSeqRecs due to communication budget limitations and on-device computational resource constraints, our \modelname employs a pretrained LLM server during the original FedSeqRec training process. Specifically, \modelname utilizes two embedding services provided by the LLM server: item-level embedding and sequence-level embedding. With the assistance of these advanced embeddings, a lightweight sequential model can achieve competitive performance. Notably, as shown in Figure~\ref{fig_arch}, \modelname is compatible with most FedSeqRecs, as it introduces the LLM model as a service (i.e., the green dashed line box part) without making any additional assumptions about the basic FedSeqRecs protocol. In the following two subsections, we will present the technical details of incorporating these two services, respectively.
Algorithm~\ref{alg_fellas} describes \modelname in pseudo-code.

\begin{algorithm}[!ht]
  \renewcommand{\algorithmicrequire}{\textbf{Input:}}
  \renewcommand{\algorithmicensure}{\textbf{Output:}}
  \caption{The pseudo-code for \modelname.} \label{alg_fellas}
  \begin{algorithmic}[1]
    \Require global round $R$; learning rate $lr$, \dots
    \Ensure  well-trained sequential model $\mathbf{M}^{R}$
    \State server initializes model $\mathbf{M}^{0}$
    \State // the central server invokes LLM item-level embedding service
    \State $\mathbf{V}^{llm}\leftarrow$\Call{LLMItemService}{$\mathcal{V}^{tx}$}
    \For {each round r =0, ..., $R-1$}
      \State sample a fraction of clients $\mathcal{U}^{r}$ from $\mathcal{U}$
        \For{$u_{i}\in \mathcal{U}^{r}$ \textbf{in parallel}} 
        \State // execute on client sides
        \State $\mathbf{M}_{u_{i}}^{r+1}\leftarrow$\Call{ClientTrain}{$u_{i}$, $\mathbf{M}^{r}$, $\mathbf{V}^{llm}$}
        \EndFor
      \State // execute on central server
      \State $\mathbf{M}^{r+1}\leftarrow$ aggregate received client model parameters $\{\mathbf{M}_{u_{i}}^{r+1}\}_{u_{i}\in \mathcal{U}^{r}}$
      \EndFor
    \Function{ClientTrain} {$u_{i}$, $\mathbf{M}^{r}$, $\mathbf{V}^{llm}$}
    \If{query for LLM sequence-level embedding service}
      \State // generate perturbed sequence to protect the raw data
      \State $\widetilde{\mathcal{S}}_{u_{i}}\leftarrow$ execute $d_{\mathcal{X}}-$privacy protection algorithm (Algorithm~\ref{alg_dx-privacy})
      \State $\mathcal{S}_{rd}\leftarrow$ generate some random sequence
      \State $\mathbf{s}_{u_{i}}^{llm}, \mathbf{s}_{rd}^{llm}\leftarrow$\Call{LLMSequenceService}{$\widetilde{\mathcal{S}}_{u_{i}}\cup \mathcal{S}_{rd}$}
    \EndIf
    \State $\mathbf{M}_{u_i}^{r+1}\leftarrow$ update local model with objective $\mathcal{L}^{rec} + \alpha\mathcal{L}^{cl}$
    \State \Return $\mathbf{M}_{u_{i}}^{r+1}$
    \EndFunction
    \end{algorithmic}
\end{algorithm}

\subsection{Item-level Embedding Service}\label{sec_item_embedding_service}
In centralized recommendation systems, numerous studies~\cite{tran2022aligning,guo2020leveraging} have demonstrated the effectiveness of leveraging textual information (e.g., item titles) to enhance ID-based item embeddings. Inspired by these successful practices, \modelname also utilizes item textual information to boost the performance of sequential models, as shown in the upper green dashed box of Figure~\ref{fig_arch}.

Specifically, at the initial stage, the central server sends all items' textual information $\mathcal{V}^{tx}=\{v_{1}^{tx},v_{2}^{tx},\dots,v_{\left|\mathcal{V}\right|}^{tx}\}$ to the LLM server and invokes the item embedding service:
\begin{equation}
  \mathbf{V}^{llm} = \mathcal{F}^{llm}(\mathcal{V}^{tx})
\end{equation}
where $\mathcal{F}^{llm}$ is the LLM-based item encoder and $\mathbf{V}^{llm}$ is its encoding output.
Then, \modelname incorporates an adapter layer $\phi$ to transform $\mathbf{V}^{llm}$ to the same dimension of the ID-based embedding and fuses them through summation:
\begin{equation}
  \mathbf{V} \leftarrow \mathbf{\phi}(\mathbf{V}^{tx}) + \mathbf{V}
\end{equation}
Incorporating the item-level embedding service in FedSeqRecs does not raise privacy concerns, as only item meta-information is delivered, which is publicly accessible on most platforms.

\subsection{Privacy-preserving Sequence-level Embedding Service}
Simply utilizing LLM-encoded item textual embeddings does not fully exploit the LLM's powerful language understanding capabilities. 
In \modelname, we also leverage LLMs to encode user interaction sequences, providing sequence-level textual information to assist the lightweight sequential recommendation model in mining user preferences. 
However, achieving such a service in FedSeqRec is challenging because it requires clients to send their interaction sequences to the LLM server for encoding, which is strictly prohibited in FedSeqRec to protect user privacy.

To incorporate this service, we employ a $d_{\mathcal{X}}$-privacy-compliant sequence perturbation method.
As depicted in the lower part of Figure~\ref{fig_arch}, before the client $u_{i}$ queries the sequence representation service, it first applies the perturbation method to its original sequence $\mathcal{S}_{u_{i}}$ obtaining a noisy sequence $\widetilde{\mathcal{S}}_{u_{i}}$.
This noisy sequence provides privacy guarantees while still maintaining a significant degree of data utility.
Next, client $u_{i}$ uploads $\widetilde{\mathcal{S}}_{u_{i}}$ to the external LLM server and obtains the encoded sequence representation $\mathbf{s}_{u_{i}}^{llm}$.
Since $\mathbf{s}_{u_{i}}^{llm}$ is derived from the noisy sequence, directly incorporating it into the local model, as with item textual embeddings, does not yield satisfactory results. 
Therefore, we propose a contrastive enhancement learning method to effectively transfer the useful information from $\mathbf{s}_{u_{i}}^{llm}$ to the sequential model.

\begin{algorithm}[!ht]
  \renewcommand{\algorithmicrequire}{\textbf{Input:}}
  \renewcommand{\algorithmicensure}{\textbf{Output:}}
  \caption{The pseudo-code for $d_{\mathcal{X}}-$privacy implementation.} \label{alg_dx-privacy}
  \begin{algorithmic}[1]
    \Require user interaction sequence $\mathcal{S}$, $\mathbf{V}^{llm}$ privacy parameter $\epsilon$
    \Ensure  perturbed sequence $\widetilde{\mathcal{S}}$
      \State $\widetilde{\mathcal{S}}=\emptyset$
      \For{$v_{t}\in \mathcal{S}$} 
        \State $\mathbf{z}\leftarrow$ sample a noise vector with density $exp(-\epsilon \left\|\mathbf{z}\right\|)$
        \State $\bar{\mathbf{v}}_{llm,t} \leftarrow \mathbf{v}_{llm,t} + \mathbf{z}$
        \State $\widetilde{v}_{t}\leftarrow \mathop{argmax}\nolimits_{v\in \mathcal{V}} sim(\mathbf{v}_{llm}, \bar{\mathbf{v}}_{llm,t})$ 
        \State add $\widetilde{v}_{t}$ into $\widetilde{\mathcal{S}}$
      \EndFor
    \end{algorithmic}
\end{algorithm}

\subsubsection{$d_{\mathcal{X}}$-privacy based Sequence Perturbation.}
$d_{\mathcal{X}}$-privacy~\cite{chatzikokolakis2013broadening} is an extended form of local differential privacy that has been widely used in textual content privacy protection~\cite{feyisetan2020privacy,qu2021natural}, as it largely preserves the original semantics of the protected content.

\textit{Definition of $d_{\mathcal{X}}$-privacy.}  
Let $\mathcal{A}: \mathcal{X}\rightarrow \mathcal{Y}$ be a randomized algorithm where $\mathcal{X}$ is its input domain and $\mathcal{Y}$ is the output range.
$\mathcal{A}$ is said to provide $\epsilon d_{\mathcal{X}}$-privacy if and only if for any two input data $x, \widetilde{x} \in \mathcal{X}$ and for any possible output $y\in \mathcal{Y}$ satisfied:
\begin{equation}
  \frac{Pr(\mathcal{A}(x) = y)}{Pr(\mathcal{A}(\widetilde{x}) = y)} \leq exp(\epsilon d_{\mathcal{X}}(x,\widetilde{x}))  
\end{equation}
where $d_{\mathcal{X}}$ is a distance function defined on $\mathcal{X}$ and $\epsilon$ is the privacy parameter that controls the privacy protection strength.
Note that same as differential privacy, $d_{\mathcal{X}}-$privacy also has the post-processing property~\cite{chatzikokolakis2013broadening}.
That is to say, if $\mathcal{A}$ satisfied $d_{\mathcal{X}}-$privacy, then for any deterministic or randomized algorithm $\mathcal{A}'$, $\mathcal{A}\circ \mathcal{A}'$ satisfies $d_{\mathcal{X}}-$privacy.

\textit{Implementation of $d_{\mathcal{X}}$-privacy.} In \modelname, users' interacted items are the most important privacy protection objective, therefore, we replace each item in user sequence using $d_{\mathcal{X}}$-privacy mechanism.
Specifically, for user $u_{i}$'s interaction sequence $\mathcal{S}_{u_{i}}=[v_{1}^{u_{i}}, \dots, v_{T_{u_{i}}}^{u_{i}}]$, we first utilize the textual item embedding $\mathbf{V}^{llm}$ obtained in Section~\ref{sec_item_embedding_service} to map them into a sequence of vectors $[\mathbf{v}_{llm,1}^{u_{i}}, \dots, \mathbf{v}_{llm,T_{u_{i}}}^{u_{i}}]$.
Then, for item $\mathbf{v}_{llm,t}^{u_{i}}$ in $\mathcal{S}_{u_{i}}$, we add noise $\mathbf{z}$ to item's embedding vector:
\begin{equation}
  \bar{\mathbf{v}}_{llm,t}^{u_{i}} = \mathbf{v}_{llm,t}^{u_{i}} + \mathbf{z}
\end{equation}
where $\mathbf{z}$ is sampled from $N$-dimension distribution with density $exp(-\epsilon \left\|\mathbf{z}\right\|)$. $N$ is the dimension of $\mathbf{v}_{llm}$.
After that, we replace the interacted item $v_{t}^{u_{i}}$ with $\widetilde{v}_{t}^{u_{i}}$, which has the smallest distance (i.e., most similar) to its perturbed embedding vector.
In this paper, we utilize the widely used cosine similarity (denoted as $sim(\cdot)$) as the measure function:
\begin{equation}
  \widetilde{v}_{t}^{u_{i}} = \mathop{argmax}\nolimits_{v\in \mathcal{V}} sim(\mathbf{v}_{llm}, \bar{\mathbf{v}}_{llm,t}^{u_{i}})
\end{equation}
Algorithm~\ref{alg_dx-privacy} briefly summarizes the implementation of $d_{\mathcal{X}}-$privacy perturbation.

\textit{Theorem. Denote the above implementation as mechanism $\mathcal{A}: \mathcal{S} \rightarrow \widetilde{\mathcal{S}}$, let metric $d(v, \widetilde{v})=\sum_{t=1}^{T} sim(\mathbf{v}_{llm,t}, \widetilde{\mathbf{v}}_{llm,t})$, then, $\mathcal{A}$ satisfies $\epsilon d_{\mathcal{X}}$-privacy with respect to the metric $d$.}

\textit{Proof.} As $\mathcal{A}$ can be viewed as a combination of the Laplace implemented exponential mechanism construction for the metric $d$ with some post-processing that does not affect privacy guarantee, $\mathcal{A}$ satisfies $\epsilon d_{\mathcal{X}}$-privacy.

\subsubsection{Contrastive Enhancement} 
When the LLM server receives the noisy sequence $\widetilde{\mathcal{S}}_{u_{i}}$, it returns the sequence's vector $\mathbf{s}_{u_{i}}^{llm}$ to the client.
To effectively leverage this vector, we design a contrastive learning method to align the information from the sequential recommendation's sequence vector $\mathbf{s}_{u_{i}}$ with $\mathbf{s}_{u_{i}}^{llm}$.
The motivation is that, since $\widetilde{\mathcal{S}}_{u_{i}}$ is generated from $\mathcal{S}_{u_{i}}$, $\mathbf{s}_{u_{i}}^{llm}$ can be treated as a contrastive view:
\begin{equation}\label{eq_contrastive}
  \mathcal{L}^{cl} = -log \frac{exp(sim(\mathbf{s}_{u_{i}},\psi(\mathbf{s}_{u_{i}}^{llm})))}{exp(sim(\mathbf{s}_{u_{i}},\psi(\mathbf{s}_{u_{i}}^{llm}))) + \sum_{s_{j}\in \mathcal{S}_{rd}} exp(sim(\mathbf{s}_{u_{i}},\psi(\mathbf{s}_{j}^{llm})))}
\end{equation}
where $\psi$ is a transformation layer that converts the LLM's encoded embedding to the same dimension as the sequential recommendation's vectors.
$\mathcal{S}_{rd}$ is a randomly formed sequence set.
In \modelname, the set size is set to 1 for efficiency.
As a result, the final local training objective for each client is changed to:
\begin{equation}\label{eq_final_loss}
  \mathcal{L} = \mathcal{L}^{rec} + \alpha\mathcal{L}^{cl}
\end{equation}
where $\alpha$ controls the contributions of contrastive enhancement. 

\begin{figure}[!t]
  \centering 
  \includegraphics[width=0.4\linewidth]{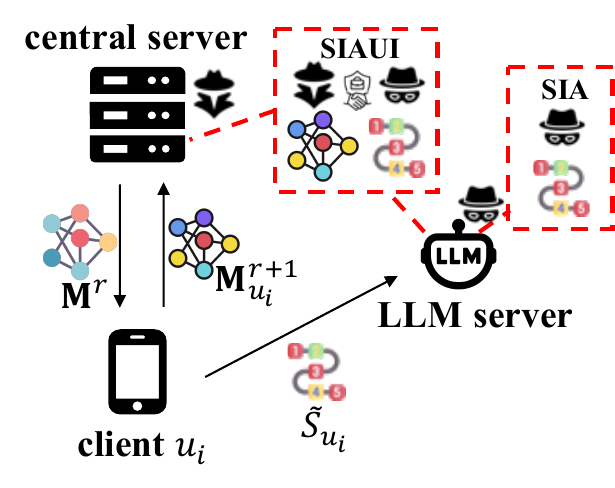}\caption{The illustration of two attacks under different scenarios.}
  \label{fig_attack}
  \end{figure}

\subsection{Interacted Item Inference Attacks}
In \modelname, clients send sequences to an external LLM server, raising additional privacy concerns compared to vanilla FedSeqRec. 
Although the sequences are perturbed with $d_{\mathcal{X}}$-privacy guarantees, to fully address users' privacy concerns, we further design two inference attacks based on users' uploaded noisy sequences to verify the empirical privacy risks of data leakage.
As illustrated in Figure~\ref{fig_attack}, these two attacks are launched in the most practical adversarial settings: (1) the external LLM server acting as a curious-but-honest server, and (2) both the LLM server and the central server being curious-but-honest and cooperating maliciously. 
Here,  ``curious-but-honest'' means that the server is interested in inferring users' interacted items but will only utilize its legally received information.
Furthermore, we assume that the attacker already knows the client's distance function in $d_{\mathcal{X}}$-privacy protection, presenting an even harsher scenario for privacy protection. 

\subsubsection{Similarity-based Inference Attack (SIA)}
In this attack, the LLM server utilizes the only prior knowledge $\widetilde{\mathcal{S}}_{u_{i}}$, to guess client $u_{i}$'s interacted items.
Specifically, for an item $\widetilde{v}_{t}^{u_{i}} \in \widetilde{\mathcal{S}}_{u_{i}}$, the LLM server first converts it into a textual item embedding $\widetilde{\mathbf{v}}_{llm,t}^{u_{i}}$.
Then, the LLM server guesses the interacted item as the one most similar to $\widetilde{v}_{t}^{u_{i}}$ from the whole item set $\mathcal{V}$ evaluated by the distance function:
\begin{equation}
  \dot{v}_{t}^{u_{i}} = \mathop{argmax}\nolimits_{v\in \mathcal{V}} sim(\mathbf{v}_{llm}, \widetilde{\mathbf{v}}_{llm,t}^{u_{i}})
\end{equation}
where $\dot{v}_{t}^{u_{i}}$ is the item that SIA infers as the interacted items for $u_{i}$.

\subsubsection{Similarity-based Inference Attack with Update Information (SIAUI)}
SIAUI builds on SIA by assuming that the central server and the LLM server share a mutual interest in obtaining user $u_{i}$'s interacted items, leading them to share their prior knowledge: the user's perturbed sequence $\widetilde{\mathcal{S}}_{u_{i}}$ and the user's uploaded model $\mathbf{M}_{u_{i}}^{r+1}$.
In detail, the attacker compares $\mathbf{M}_{u_{i}}^{r+1}$ with $\mathbf{M}^{r}$ to infer a set of client $u_{i}$'s trained items $\mathcal{V}_{u_{i}}^{trained}$ by calculating the non-zero item embedding gradients. 
This trained set includes both interacted items and negative samples.
Similar to SIA, for $\widetilde{v}_{t}^{u_{i}} \in \widetilde{\mathcal{S}}_{u_{i}}$, SIAUI converts it into $\widetilde{\mathbf{v}}_{llm,t}^{u_{i}}$.
Then, SIAUI only needs to select the most similar items from the trained item set $\mathcal{V}_{u_{i}}^{trained}$, which is much easier than SIA guesses from the whole item set.
\begin{equation}\label{eq_SIAUI}
  \dot{v}_{t}^{u_{i}} = \mathop{argmax}\nolimits_{v\in \mathcal{V}_{u_{i}}^{trained}} sim(\mathbf{v}_{llm}, \widetilde{\mathbf{v}}_{llm,t}^{u_{i}})
\end{equation}
Algorithm~\ref{alg_siaui} introduces SIAUI briefly with pseudo-code.

\begin{algorithm}[!ht]
  \renewcommand{\algorithmicrequire}{\textbf{Input:}}
  \renewcommand{\algorithmicensure}{\textbf{Output:}}
  \caption{The pseudo-code for SIAUI.} \label{alg_siaui}
  \begin{algorithmic}[1]
    \Require user perturbed interaction sequence $\widetilde{\mathcal{S}}_{u_{i}}$, user uploaded parameters $\mathbf{M}_{u_{i}}^{r+1}$, the global model $\mathbf{M}^{r}$, LLM-based item embedding table $\mathbf{V}^{llm}$
    \Ensure  inferred user $u_{i}$ sequence $\dot{\mathcal{S}}_{u_{i}}$
      \State $\dot{\mathcal{S}}_{u_{i}}=\emptyset$
      \State $\mathcal{V}_{u_{i}}^{trained}\leftarrow$ select trained items according to the changes of item embeddings in $\mathbf{M}_{u_{i}}^{r+1}$ and $\mathbf{M}^{r}$
      \For{$\widetilde{v}_{t}\in \widetilde{\mathcal{S}}_{u_{i}}$} 
        \State $\dot{v}_{t}^{u_{i}}\leftarrow$ replace $\widetilde{v}_{t}$ using E.q.~\ref{eq_SIAUI}
        \State add $\{\dot{v}_{t}^{u_{i}}\}$ into  $\dot{\mathcal{S}}_{u_{i}}$
        \State remove $\{\dot{v}_{t}^{u_{i}}\}$ from the ``guess pool'' $\mathcal{V}_{u_{i}}^{trained}$ 
      \EndFor
    \end{algorithmic}
\end{algorithm}

\subsection{Discussion}
\subsubsection{Privacy Protection Discussion}
In this work, for clarity, we focus primarily on privacy protection when querying external LLM services.
This protection is orthogonal with the privacy protection in FedSeqRec which has been widely researched~\cite{yuan2023interaction}.
In other words, \modelname is compatible with most existing privacy protection methods designed for FedSeqRec systems.

\subsubsection{Communication Cost Discussion}
Compared to traditional FedSeqRec, the primary additional communication cost comes from calling the LLM services. Specifically, for the central server, the extra cost involves sending the textual information $\mathcal{V}^{tx}$ to the LLM server, but this is a one-time expense at the beginning of the training process. For the clients, the additional cost comes from transmitting the noisy sequence $\widetilde{\mathcal{S}}$, which is just a sequence of numbers. Given the sparse nature of interaction data, this communication cost is minimal.

\section{Experiments}\label{sec_experiments}
In this section, we first introduce the experimental settings. Then, we report and analyze the experimental results to answer the following research questions:
\begin{itemize}
  \item \textbf{RQ1.} Can \modelname effectively and efficiently improve the performance of FedSeqRec?
  \item \textbf{RQ2.} Is the perturbation module effective in protecting users' private data?
  \item \textbf{RQ3.} Do the two services contribute to the improvement of \modelname?
  \item \textbf{RQ4.} What are the impacts of the contrastive learning factor?
  \item \textbf{RQ5.} Further Analysis: What are the impacts of different LLM models?
\end{itemize}

\subsection{Datasets}
We conduct experiments on three widely used datasets collected from the Amazon platform that cover different recommendation domains (Beauty, Office Product (Office) and Patio Lawn and Garden (PLG))~\cite{ni2019justifying} to validate the effectiveness of \modelname.
The statistics of these three datasets are in Table~\ref{tb_dataset}.
Beauty includes $20,574$ users and $10,020$ beauty-related products with $164,268$ interactions.
Office consists of $42,268$ reviews among $4,619$ users and $1,939$ office products, while PLG contains $1,518$ users with $10,439$ purchases history on $751$ patio, lawn and garden-related items.
Following the most common data preprocessing procedure~\cite{jannach2017recurrent,zhou2020s3,lei2021semi,kang2018self} in the sequential recommendation, we transform all review ratings to implicit feedback (i.e., $r=1$) and sorted them according to the interacted timestamp to form user interaction sequence.
The maximum interaction sequence length is set to $50$.
Then, the last two items in each sequence are used for validation and testing while the remains are for training.

\begin{table}[!htbp]
  \centering
  \caption{Statistics of three datasets used in our experiments.}\label{tb_dataset}
  \begin{tabular}{l|cccc}
  \hline
  \textbf{Dataset}        & \textbf{Beauty} & \textbf{Office} & \textbf{PLG}  \\ \hline
  \textbf{\#Users}        & 20,574      & 4,619        & 1,518          \\
  \textbf{\#Items}        & 10,020       & 1,939         & 751           \\
  \textbf{\#Interactions} & 164,268     & 42,268        & 10,439        \\
  \textbf{Avgerage Lengths} & 7.9      & 9.1         & 6.8          \\
  \textbf{Density} & 0.07\%             & 0.47\%        & 0.91\%       \\ \hline
  \end{tabular}
  \end{table}

  \begin{table}[!htbp]
    \centering
    \caption{The recommendation performance of \modelname and baselines on three datasets. ``Cent.'' is short for centralized settings, ``Fed.'' is abbreviated for FedSeqRec and ``ours'' is for \modelname. ``Stand. LLM'' means directly using the LLM for recommendation.}\label{tb_main}
    \begin{tabular}{ll|c|ccc|ccc}
    \hline
    \textbf{}                                             & \textbf{}     & \textbf{}             & \multicolumn{3}{c|}{\textbf{GRU4Rec}}               & \multicolumn{3}{c}{\textbf{SASRec}}                 \\ \cline{4-9} 
    \textbf{}                                             & \textbf{}     & \textbf{Stand. LLM} & \textbf{Cent.}   & \textbf{Fed.} & \textbf{Ours}    & \textbf{Cent.}   & \textbf{Fed.} & \textbf{Ours}    \\ \hline
    \multicolumn{1}{l|}{\multirow{4}{*}{\textbf{Beauty}}} & \textbf{H@10} & 0.00111               & {\ul 0.02201}    & 0.01764       & \textbf{0.02454} & {\ul 0.02289}    & 0.01769       & \textbf{0.02532} \\
    \multicolumn{1}{l|}{}                                 & \textbf{N@10} & 0.00039               & {\ul 0.01130}    & 0.00827       & \textbf{0.01230} & {\ul 0.01133}    & 0.00839       & \textbf{0.01264} \\
    \multicolumn{1}{l|}{}                                 & \textbf{H@20} & 0.00155               & {\ul 0.03455}    & 0.02751       & \textbf{0.03465} & {\ul 0.03494}    & 0.03052       & \textbf{0.04009} \\
    \multicolumn{1}{l|}{}                                 & \textbf{N@20} & 0.00050               & {\ul 0.01410}    & 0.01073       & \textbf{0.01410} & {\ul 0.01439}    & 0.01088       & \textbf{0.01633} \\ \hline
    \multicolumn{1}{l|}{\multirow{4}{*}{\textbf{Office}}} & \textbf{H@10} & 0.00432               & {\ul 0.02208}    & 0.01407       & \textbf{0.02944} & {\ul 0.01342}    & 0.00757       & \textbf{0.01623} \\
    \multicolumn{1}{l|}{}                                 & \textbf{N@10} & 0.00185               & {\ul 0.00970}    & 0.00678       & \textbf{0.01458} & {\ul 0.00605}    & 0.00325       & \textbf{0.00750} \\
    \multicolumn{1}{l|}{}                                 & \textbf{H@20} & 0.00757               & {\ul 0.03875}    & 0.02836       & \textbf{0.05477} & {\ul 0.02987}    & 0.01623       & \textbf{0.03810} \\
    \multicolumn{1}{l|}{}                                 & \textbf{N@20} & 0.00264               & {\ul 0.01390}    & 0.01037       & \textbf{0.02095} & {\ul 0.01010}    & 0.00545       & \textbf{0.01290} \\ \hline
    \multicolumn{1}{l|}{\multirow{4}{*}{\textbf{PLG}}}    & \textbf{H@10} & 0.00790               & \textbf{0.03030} & 0.02239       & {\ul 0.02700}    & {\ul 0.03293}    & 0.03096       & \textbf{0.03359} \\
    \multicolumn{1}{l|}{}                                 & \textbf{N@10} & 0.00418               & \textbf{0.01337} & 0.01076       & {\ul 0.01324}    & {\ul 0.01529}    & 0.01432       & \textbf{0.01601} \\
    \multicolumn{1}{l|}{}                                 & \textbf{H@20} & 0.01646               & \textbf{0.05204} & 0.04018       & {\ul 0.04874}    & \textbf{0.05994} & 0.05401       & 0.05665          \\
    \multicolumn{1}{l|}{}                                 & \textbf{N@20} & 0.00630               & \textbf{0.01917} & 0.01480       & {\ul 0.01867}    & {\ul 0.02152}    & 0.02021       & \textbf{0.02182} \\ \hline
    \end{tabular}
    \end{table}

    \begin{figure}[!htbp]
      \centering
      \includegraphics[width=1.\textwidth]{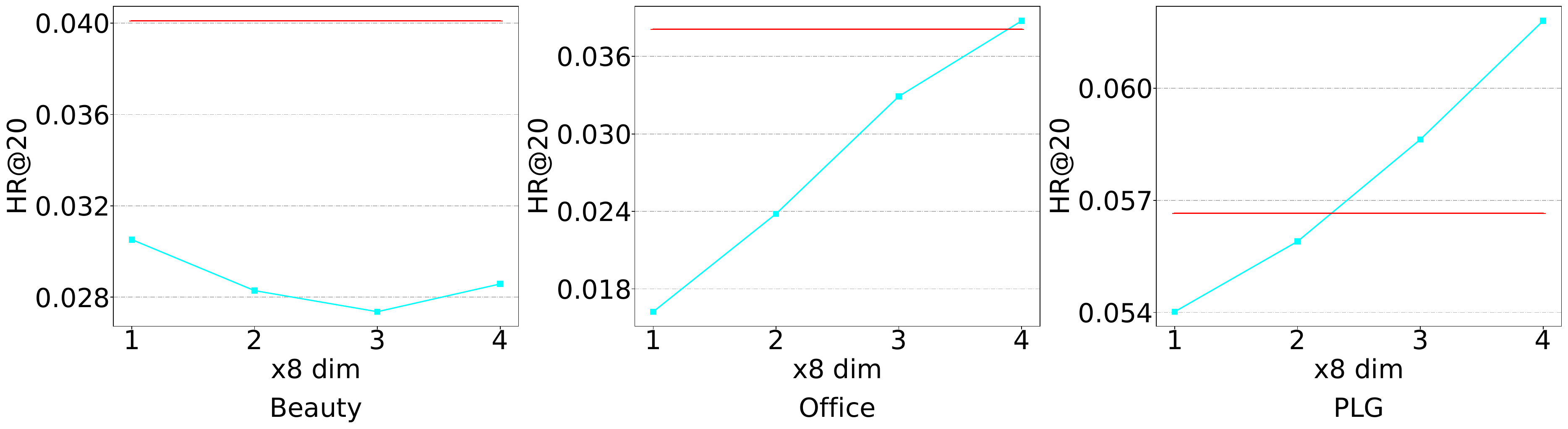}
      \caption{The performance of lightweight \modelname (\textcolor{red}{red line}) and vanilla FedSeqRec (\textcolor{cyan}{cyan line}) with increased dimension sizes.}\label{fig_dimension}
    \end{figure}

\subsection{Baselines}
As \modelname is a generic framework based on the vanilla federated protocol and compatible with most sequential recommendation models, we utilize the most commonly used techniques, GRU4Rec~\cite{chung2014empirical} and SASRec~\cite{kang2018self}, as the base models. We then compare these with their \emph{centralized} and vanilla \emph{federated} versions to assess the recommendation improvements. Additionally, we compare using solely the LLM server to predict users' preferences without federated sequential recommendation training, to highlight the necessity of training sequential recommendation models.

For evaluating empirical privacy protection abilities, we compare \modelname's perturbation method with random perturbation, where items in user sequences are randomly replaced with other items.

\subsection{Evaluation Protocol}
We evaluate the model performance using two widely used metrics: Hit Ratio at Rank K (HR@K) and Normalized Discounted Cumulative Gain at Rank K (NDCG@K). HR measures the ratio of correct items included in the top K recommendation list, while NDCG further evaluates whether the correct items are positioned high on the list. In this paper, K is set to 10 and 20. We calculate the metric scores for the entire item set that has not been interacted with by users to avoid evaluation bias~\cite{krichene2020sampled}. 

For the privacy protection assessment, we use the F1 scores of attacks as the metrics.

\subsection{Implementation Details}
For both GRU4Rec and SASRec, we set the item embedding dimension and hidden vector sizes to $8$ in \modelname and compare these lightweight base models with larger models (up to $4$ times the size) in the vanilla FedSeqRec to demonstrate the efficiency of our proposed methods.
$\phi$ and $\psi$ are single-layer MLPs whose input size matches the LLM's encoding vector, and the output size matches the base recommendation's embedding dimension.
We employ Adam with a $0.001$ learning rate as the optimizer.
$\frac{1}{\epsilon}$ is set to $0.01$ as default and we evaluate its impact in Section~\ref{sec_privacy_exp}.
$\alpha$ is searched in $[1.0, 0.1, 0.01, 0.001]$ to find the best performance.
The maximum global round is $20$.
We define a whole global round as having traversed all clients in the training process. 
Specifically, before a global round begins, we shuffle the client queue and then traverse the queue by selecting $256$ clients at a time to participate in the training process. 
For each client, the local training epochs are set to $5$, and the negative sampling ratio is set to $1:1$. 

Due to computational power limitations, we utilize Longformer\footnote{\url{https://huggingface.co/allenai/longformer-base-4096}}~\cite{beltagy2020longformer} as the LLM server by default, considering its computing efficiency in handling long context input. 
It is worth noting that \modelname does not have constraints on LLM types. 
To demonstrate its generalization on LLM servers, we also explore other different LLMs, including classic LLM (BERT\footnote{\url{https://huggingface.co/google-bert/bert-base-cased}}~\cite{devlin2019bert}) and cutting-edge techniques (Llama3.1\footnote{\url{https://huggingface.co/meta-llama/Meta-Llama-3-8B-Instruct}} and Llama2\footnote{\url{https://huggingface.co/meta-llama/Llama-2-7b-chat-hf}}~\cite{touvron2023llama}), in Section~\ref{sec_llm_exp}.
We utilize item titles as the main input for LLMs. For the sequence representation service, we use the prefix ``The user's purchase history list is as follows:'' as the prompt. 
More complex prompts and additional item meta-information can be explored in future research. The last layer's hidden vector of the start special token is used as the embedding vector.

Note that when investigating RQ2 to RQ5, we utilize SASRec as the base model by default, and we use \modelname or FedSeqRec to directly refer to \modelnamenospace(SASRec) or FedSeqRec(SASRec) respectively for concise presentation.

\begin{table*}[!htbp]
  \centering
  \caption{The attack performance (F1 scores) for different perturbation settings on three datasets. Lower values indicate better data privacy protection abilities.}\label{tb_attack}
  \begin{tabular}{l|cc|cc|cc}
  \hline
                                                & \multicolumn{2}{c|}{\textbf{Beauty}} & \multicolumn{2}{c|}{\textbf{Office}} & \multicolumn{2}{c}{\textbf{PLG}} \\ \hline
                                                & \textbf{SIA}     & \textbf{SIAUI}    & \textbf{SIA}     & \textbf{SIAUI}    & \textbf{SIA}   & \textbf{SIAUI}  \\ \hline
  \textbf{Random}                               & 0.0              & 0.52335           & 0.00386          & 0.48229           & 0.00805        & 0.50155         \\
  \textbf{\modelnamenospace($\frac{1}{\epsilon}$=0.1)}   & 0.0              & 0.54240           & 0.00064          & 0.53989           & 0.00222        & 0.47361         \\
  \textbf{\modelnamenospace($\frac{1}{\epsilon}$=0.01)}  & 0.0              & 0.55754           & 0.00534          & 0.58005           & 0.02830        & 0.56246         \\
  \textbf{\modelnamenospace($\frac{1}{\epsilon}$=0.001)} & 0.22998          & 0.90565           & 0.26651          & 0.88023           & 0.24005        & 0.87870         \\ \hline
  \end{tabular}
  \end{table*}

\begin{figure}[!htbp]
  \centering
  \includegraphics[width=1.\textwidth]{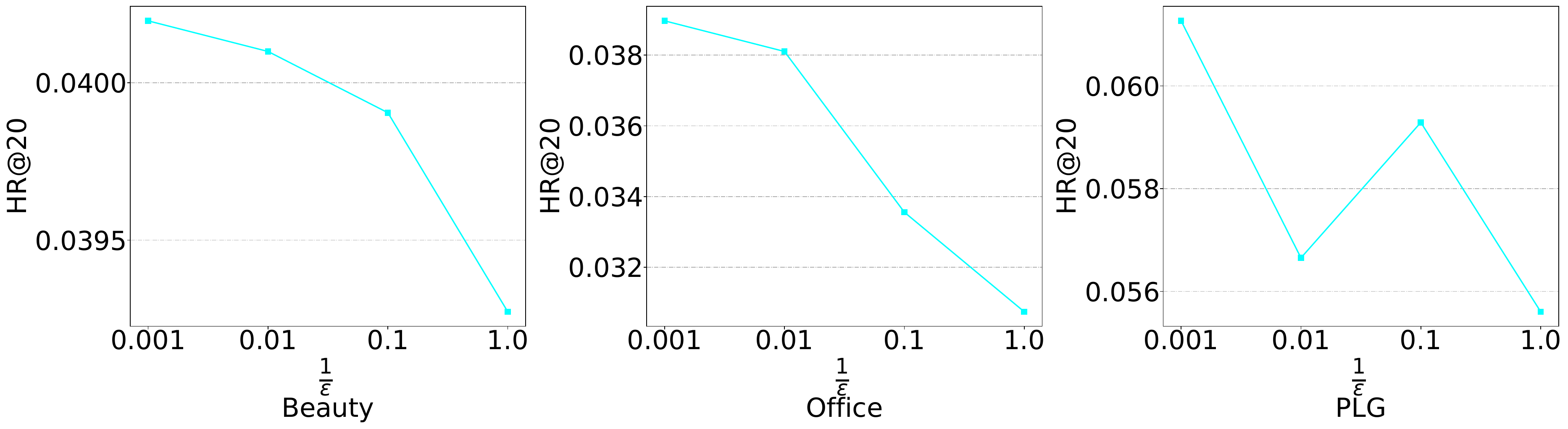}
  \caption{The recommendation performance trend (HR@20) of \modelname  with different privacy parameter $\epsilon$ on three datasets.}\label{fig_privacy_factor}
\end{figure}

\subsection{RQ1. Performance Improvement}
The core idea of \modelname is to leverage LLM services to improve the representation ability of lightweight sequential recommendation models, thereby enhancing FedSeqRec's recommendation performance. Therefore, we first present the recommendation performance using different training frameworks (i.e., centralized, FedSeqRec, and \modelname) in Table~\ref{tb_main}. 

The following observations can be made according to the empirical results:
(1) By comparing vanilla FedSeqRec with centralized training, we observe a significant performance gap. This may be because, in FedSeqRec, each client has an extremely limited training corpus (each client only has one interaction sequence), and the model updates on the local training resource are far from optimal.
(2) When equipped with \modelname, FedSeqRec shows significant performance improvements in all cases. Moreover, in most cases (e.g., on Beauty and Office datasets), \modelname even outperforms the centralized counterparts, demonstrating the effectiveness of enhancing FedSeqRec.
(3) Directly leveraging LLMs to recommend items to users results in poor performance, indicating the necessity of FedSeqRec training.

To further demonstrate the effectiveness of \modelname, we compare our method with another straightforward way of improving sequential recommendation representation ability: increasing the model size. 
Intuitively, this naive method may be effective when the model is lightweight and has poor representation abilities, but it will also significantly increase the communication and computation burden.
In Figure~\ref{fig_dimension}, we compare \modelname with vanilla FedSeqRec, increasing its base model's sizes from $1$ to $4$ times larger than \modelname's.
As shown in the figure, on Beauty, the FedSeqRec's performance does not improve as the model size increases. This is because the dataset is extremely sparse, as implied in Table~\ref{tb_dataset}, making it difficult for the larger model to converge.
Additionally, in Office and PLG, increasing model size can improve final performance. 
\modelname achieves comparable performance with the four-times-larger model trained by FedSeqRec on Office and surpasses the model with two times more parameters on PLG, indicating \modelname's efficiency in improving FedSeqRec performance.

\begin{table}[!htbp]
  \centering
  \caption{Ablation study (HR@20 scores) for two LLM services on three datasets.}~\label{tb_ablation}
  \begin{tabular}{l|ccc}
  \hline
                                      & \textbf{Beauty} & \textbf{Office} & \textbf{PLG} \\ \hline
  \textbf{\modelname} & 0.04009         & 0.03810         & 0.05665      \\
  \textbf{-SeqEmb}                   & 0.03766         & 0.03550         & 0.05467      \\
  \textbf{-SeqEmb-ItemEmb}           & 0.03052         & 0.01623         & 0.05401      \\ \hline
  \end{tabular}
  \end{table}

\begin{figure}[!htb]
  \centering
  \includegraphics[width=1.\textwidth]{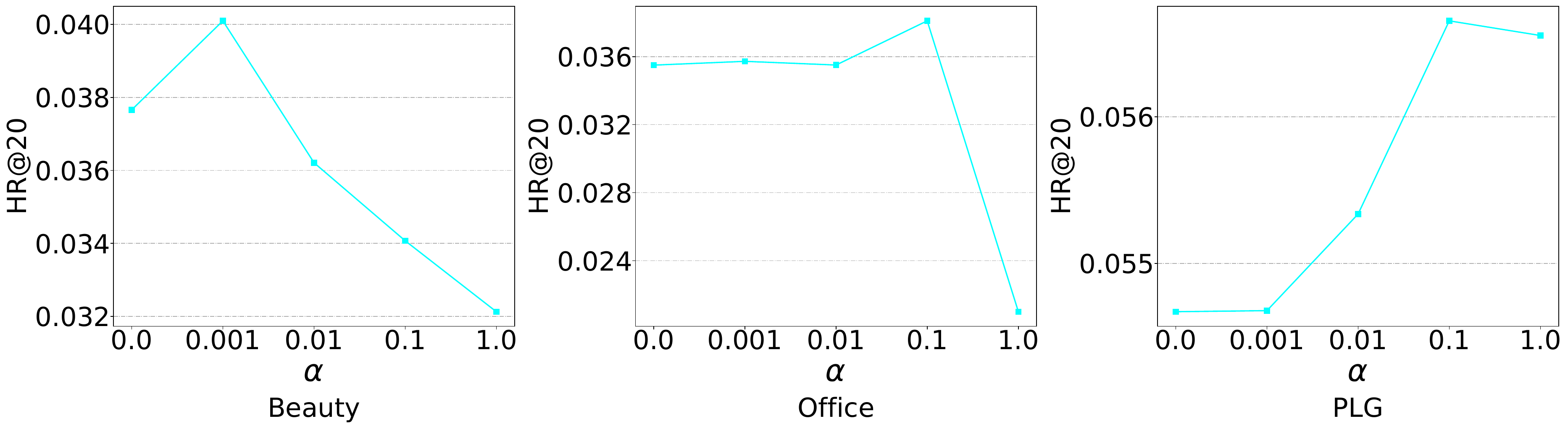}
  \caption{The recommendation performance (HR@20) trend of \modelname with different contrastive factor $\alpha$ on three datasets.}\label{fig_contrastive_factor}
\end{figure} 
 
\subsection{RQ2. Effectiveness of Privacy Protection Module}~\label{sec_privacy_exp}
For a federated sequential recommendation system, beyond recommendation performance, a critical aspect is the ability to protect user data privacy, which is the main motivation for combining federated learning with sequential recommendation models.
In this paper, we employ $d_{\mathcal{X}}$-privacy, an extension of Local Differential Privacy (LDP), to provide theoretical privacy guarantees when clients share data with external LLMs.
In this part, We further investigate two attack scenarios to empirically study the data leakage problem in \modelname. 

Table~\ref{tb_attack} presents the results of these attacks.
We include random perturbation for comparison, where items are replaced with randomly selected items.
The results show that SIAUI achieves significantly better attack outcomes than SIA due to having more prior knowledge.
When $\frac{1}{\epsilon}= 0.001$, SIAUI achieves over $0.8$ F1 scores, indicating that the privacy protection is too weak.
However, when $\frac{1}{\epsilon} \geq 0.01$, the F1 scores achieved by both attacks in \modelname are close to those in random perturbation, demonstrating strong privacy protection abilities.

In addition to defending against inference attacks, privacy protection should not significantly compromise model performance. 
Therefore, we explore the performance trend of \modelname with different privacy parameters.
As shown in Figure~\ref{fig_privacy_factor}, privacy protection strength and recommendation performance are generally negatively correlated. 
Thus, there is a trade-off between system performance and user privacy protection.
Empirically, $\frac{1}{\epsilon}=0.01$ can achieve both satisfied recommendation and privacy protection performance in this paper.

\subsection{RQ3. Ablation Study of LLM Service}~\label{sec_ablation_service}
\modelname includes two LLM services: the item-level embedding service and the sequence-level embedding service. In this subsection, we investigate the effectiveness of these two services.

According to the results on Table~\ref{tb_ablation}, removing the sequence embedding service results in HR@20 scores dropping by approximately $8\%$, $7\%$, and $4\%$ on Beauty, Office, and PLG datasets, respectively.
This demonstrates the positive impact of the sequence embedding service.
Further, when the item embedding service is also removed, the recommendation performance significantly decreases, indicating the crucial role of the item embedding service.
In conclusion, both of these services are effective in enhancing model performance.

\subsection{RQ4. The Impact of hyper-parameter $\alpha$}~\label{sec_alpha_impact}
To learn from the noisy sequence embedding $\mathbf{s}_{u_{i}}^{llm}$, clients in \modelname treat the perturbed $\widetilde{\mathcal{S}}_{u_{i}}$ as contrastive view and leverage contrastive learning to integrate the information.
As illustrated in E.q.~\ref{eq_final_loss}, $\alpha$ is the key hyper-parameter that controls the contribution of the contrastive loss $\mathcal{L}^{cl}$.
Hence, in this subsection, we explore the impacts of $\alpha$ on \modelname.

Figure~\ref{fig_contrastive_factor} shows the performance trends of \modelname with increasing values of $\alpha$ from $0.001$ to $1.0$ across three datasets.
Generally, as $\alpha$ increases, the model performance also improves until it reaches a peak, confirming the effectiveness of our contrastive enhancement.
However, beyond this peak, further increasing the value of $\alpha$ causes a decline in model performance, as the contrastive task becomes too dominant and significantly hinders the learning of the recommendation task.
For instance, on the Beauty dataset, \modelname achieves its best performance when $\alpha=0.001$, meanwhile, for the Office and PLG datasets, increasing $\alpha$ to 0.1 positively influences the recommendation model.

\begin{table}[!htbp]
  \centering
  \caption{The recommendation performance changes of \modelname with different LLMs on PLG.}\label{tb_llm}
  \begin{tabular}{l|cccc}
  \hline
  \textbf{PLG}        & \textbf{H@10} & \textbf{N@10} & \textbf{H@20} & \textbf{N@20} \\ \hline
  \textbf{None}       & 0.03096       & 0.01432       & 0.05401       & 0.02021             \\
  \textbf{BERT}       & 0.03098       & 0.01678       & 0.05533       & 0.02288             \\
  \textbf{Longformer} & 0.03359       & 0.01601       & 0.05665       & 0.02182       \\
  \textbf{Llama2}     & 0.03820       & 0.01785       & \textbf{0.06389}       & 0.02416             \\
  \textbf{Llama3.1}     & \textbf{0.04808}       & \textbf{0.02465}       & 0.06192       & \textbf{0.03161}       \\ \hline
  \end{tabular}
  \end{table}

\subsection{RQ5. The Impact of different LLM servers}~\label{sec_llm_exp}
In the above experiments, we use Longformer as the default LLM server due to computational resource limitations. However, \modelname does not impose any requirements on the type of LLM server used. 

In this part, we explore using diverse types of LLMs to provide embedding services, including the classic LLM such as BERT and the cutting-edge LLMs such as Llama2 and Llama3.1. 
We conduct experiments on the PLG dataset for computational efficiency due to its smaller size.
According to Table~\ref{tb_llm}, all LLM servers bring performance gains to the original FedSeqRec. The performance of using BERT as the server is worse than Longformer, likely because the latter can handle much longer context input, thus avoiding loss of input sequence information. Additionally, employing recent cutting-edge LLMs, such as Llama2 and Llama3, results in even more noticeable improvements in recommendation performance. Specifically, the Llama2 server brings nearly a 20\% performance gain in HR@20, while Llama3 achieves the best performance across all other metrics.
In conclusion, as shown in Table~\ref{tb_llm}, the performance gain of \modelname is generally positively related to the quality and advancement of the LLM server used.

\begin{figure}[!htbp]
  \centering
  \includegraphics[width=0.8\textwidth]{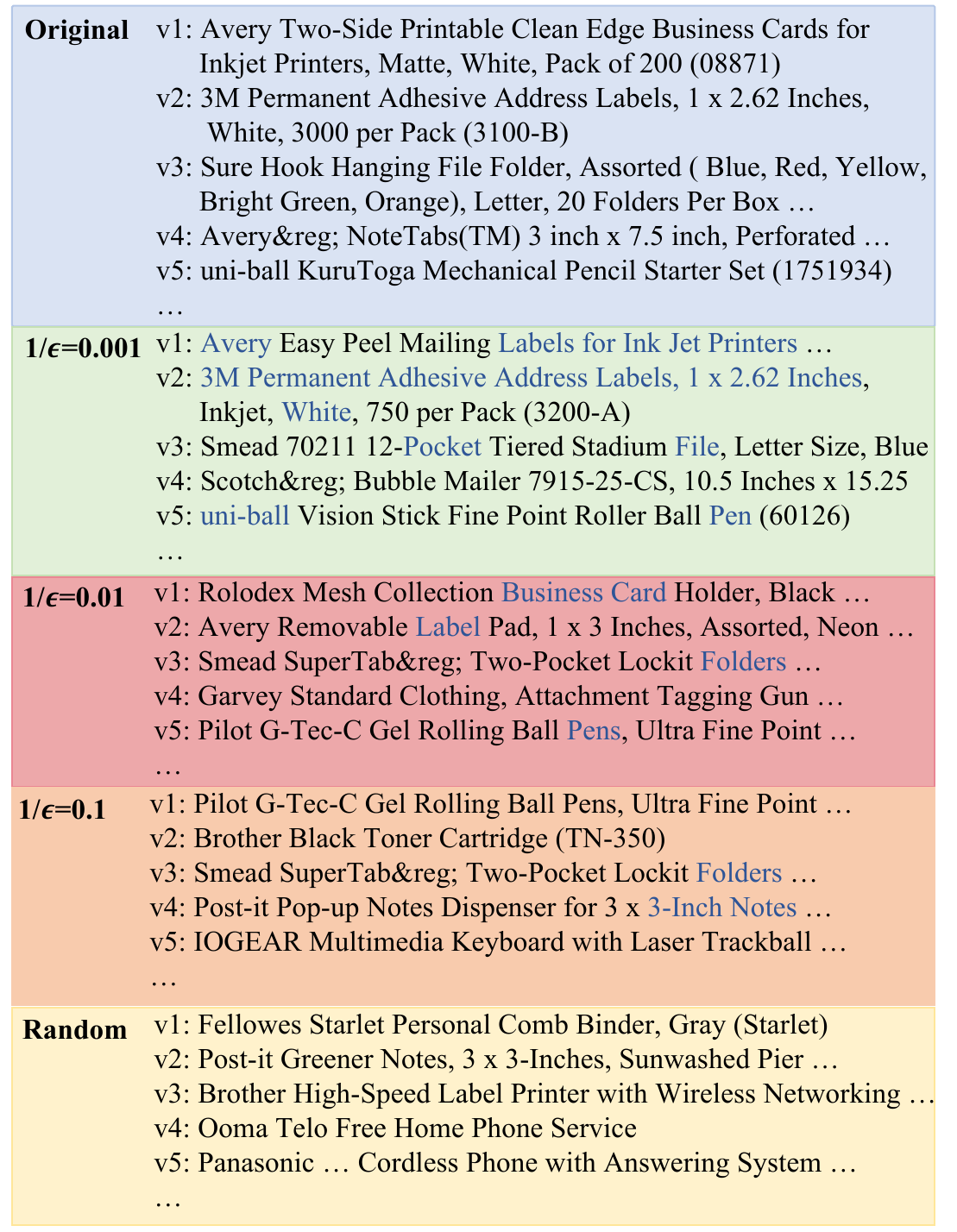}
  \caption{A case study of different sequence perturbation results. ``\textcolor{blue}{Blue text}'' highlights items with potential semantic similarity to those in the original sequence. The proposed perturbation method controls the semantic distance between the original and replaced items, enabling an adjustable trade-off between privacy and effectiveness.}\label{fig_casestudy}
\end{figure}

\subsection{Case Study for Sequence Perturbation}
To further understand the privacy protection ability of the proposed sequence perturbation methods, we randomly selected a user from the Office dataset for a case study.
For brevity, we showcase only the first five items in the user's interaction sequence, as depicted in Figure~\ref{fig_casestudy}.
As shown, the original sequence indicates that the user purchased ``cards for inkjet printers'', ``3M labels'', ``file folder'', ``notetabs'', and ``pencil set'', respectively.
When applying our $d_{\mathcal{X}}$-privacy-based protection method, all items in the original sequence are replaced, effectively safeguarding the user's data privacy. Furthermore, as the privacy budget becomes more restrictive, the replaced items tend to be semantically more distant from the original ones. This characteristic offers a flexible approach to balancing user data privacy and system effectiveness, consistent with the empirical findings in Section~\ref{sec_privacy_exp}.
Specifically, when $\frac{1}{\epsilon}=0.001$, most items are replaced with similar ones that are either from different brands or slightly different models of the same brand. However, when $\frac{1}{\epsilon}=0.1$, the replaced items retain only a few similar words in their titles. On the other hand, random replacement results in irrelevant items, offering stronger privacy protection but sacrificing all semantic meaning.

\vspace{10pt}
\section{Conclusion and Future Work}\label{sec_conclusion}
In this paper, we propose a novel framework, \modelname, to enhance federated sequential recommendation performance by incorporating advanced LLM services. Specifically, an external LLM server provides two embedding services to improve the representation abilities of lightweight sequential recommendation models: an item embedding service and a sequence embedding service.
To protect users' data privacy, we implement $d_{\mathcal{X}}$-privacy in the sequence embedding service and design a contrastive learning method to transfer sequential textual information to the recommendation model. Moreover, to empirically analyze the data leakage problem in \modelname after incorporating LLM services, we design two interacted item inference attacks under two adversarial settings.
We conduct extensive experiments on three popular recommendation datasets using two commonly used sequential recommendation models. The results and further analysis demonstrate the effectiveness and generalization of \modelname.


\emph{Future Work.} As the first work to explore the use of large language models (LLMs) as services in FedSeqRec, there are many aspects yet to be investigated. In this paper, we focus solely on using LLMs to provide representation services, leaving other promising services under-explored. For instance, in FedSeqRec, clients often struggle with data sparsity, making it difficult to train their sequential models effectively. LLMs could be used to augment clients' local datasets in a privacy-preserving manner, potentially alleviating this issue.
Additionally, this work only addresses privacy from the users' perspective, while overlooking the privacy concerns of the LLM server itself. Future research could explore how to protect the intellectual property of item embeddings and sequence embeddings generated by the LLM server.

\begin{acks}
  The Australian Research Council supports this work under the streams of Future Fellowship (Grant No. FT210100624), the Discovery Project (Grant No. DP240101108), and the Linkage Project (Grant No. LP230200892).
\end{acks}
\newpage

\bibliographystyle{ACM-Reference-Format}
\bibliography{sample-base}


\begin{thebibliography}{71}


\ifx \showCODEN    \undefined \def \showCODEN     #1{\unskip}     \fi
\ifx \showDOI      \undefined \def \showDOI       #1{#1}\fi
\ifx \showISBNx    \undefined \def \showISBNx     #1{\unskip}     \fi
\ifx \showISBNxiii \undefined \def \showISBNxiii  #1{\unskip}     \fi
\ifx \showISSN     \undefined \def \showISSN      #1{\unskip}     \fi
\ifx \showLCCN     \undefined \def \showLCCN      #1{\unskip}     \fi
\ifx \shownote     \undefined \def \shownote      #1{#1}          \fi
\ifx \showarticletitle \undefined \def \showarticletitle #1{#1}   \fi
\ifx \showURL      \undefined \def \showURL       {\relax}        \fi
\providecommand\bibfield[2]{#2}
\providecommand\bibinfo[2]{#2}
\providecommand\natexlab[1]{#1}
\providecommand\showeprint[2][]{arXiv:#2}

\bibitem[\protect\citeauthoryear{Ammad-Ud-Din, Ivannikova, Khan, Oyomno, Fu,
  Tan, and Flanagan}{Ammad-Ud-Din et~al\mbox{.}}{2019}]%
        {ammad2019federated}
\bibfield{author}{\bibinfo{person}{Muhammad Ammad-Ud-Din},
  \bibinfo{person}{Elena Ivannikova}, \bibinfo{person}{Suleiman~A Khan},
  \bibinfo{person}{Were Oyomno}, \bibinfo{person}{Qiang Fu},
  \bibinfo{person}{Kuan~Eeik Tan}, {and} \bibinfo{person}{Adrian Flanagan}.}
  \bibinfo{year}{2019}\natexlab{}.
\newblock \showarticletitle{Federated collaborative filtering for
  privacy-preserving personalized recommendation system}.
\newblock \bibinfo{journal}{\emph{arXiv preprint arXiv:1901.09888}}
  (\bibinfo{year}{2019}).
\newblock


\bibitem[\protect\citeauthoryear{Beltagy, Peters, and Cohan}{Beltagy
  et~al\mbox{.}}{2020}]%
        {beltagy2020longformer}
\bibfield{author}{\bibinfo{person}{Iz Beltagy}, \bibinfo{person}{Matthew~E
  Peters}, {and} \bibinfo{person}{Arman Cohan}.}
  \bibinfo{year}{2020}\natexlab{}.
\newblock \showarticletitle{Longformer: The long-document transformer}.
\newblock \bibinfo{journal}{\emph{arXiv preprint arXiv:2004.05150}}
  (\bibinfo{year}{2020}).
\newblock


\bibitem[\protect\citeauthoryear{Brown, Mann, Ryder, Subbiah, Kaplan, Dhariwal,
  Neelakantan, Shyam, Sastry, Askell, et~al\mbox{.}}{Brown
  et~al\mbox{.}}{2020}]%
        {brown2020language}
\bibfield{author}{\bibinfo{person}{Tom Brown}, \bibinfo{person}{Benjamin Mann},
  \bibinfo{person}{Nick Ryder}, \bibinfo{person}{Melanie Subbiah},
  \bibinfo{person}{Jared~D Kaplan}, \bibinfo{person}{Prafulla Dhariwal},
  \bibinfo{person}{Arvind Neelakantan}, \bibinfo{person}{Pranav Shyam},
  \bibinfo{person}{Girish Sastry}, \bibinfo{person}{Amanda Askell},
  {et~al\mbox{.}}} \bibinfo{year}{2020}\natexlab{}.
\newblock \showarticletitle{Language models are few-shot learners}.
\newblock \bibinfo{journal}{\emph{Advances in neural information processing
  systems}}  \bibinfo{volume}{33} (\bibinfo{year}{2020}),
  \bibinfo{pages}{1877--1901}.
\newblock


\bibitem[\protect\citeauthoryear{Chatzikokolakis, Andr{\'e}s, Bordenabe, and
  Palamidessi}{Chatzikokolakis et~al\mbox{.}}{2013}]%
        {chatzikokolakis2013broadening}
\bibfield{author}{\bibinfo{person}{Konstantinos Chatzikokolakis},
  \bibinfo{person}{Miguel~E Andr{\'e}s}, \bibinfo{person}{Nicol{\'a}s~Emilio
  Bordenabe}, {and} \bibinfo{person}{Catuscia Palamidessi}.}
  \bibinfo{year}{2013}\natexlab{}.
\newblock \showarticletitle{Broadening the scope of differential privacy using
  metrics}. In \bibinfo{booktitle}{\emph{Privacy Enhancing Technologies: 13th
  International Symposium, PETS 2013, Bloomington, IN, USA, July 10-12, 2013.
  Proceedings 13}}. Springer, \bibinfo{pages}{82--102}.
\newblock


\bibitem[\protect\citeauthoryear{Chen, Yao, McAuley, Zhou, and Wang}{Chen
  et~al\mbox{.}}{2023}]%
        {chen2023deep}
\bibfield{author}{\bibinfo{person}{Xiaocong Chen}, \bibinfo{person}{Lina Yao},
  \bibinfo{person}{Julian McAuley}, \bibinfo{person}{Guanglin Zhou}, {and}
  \bibinfo{person}{Xianzhi Wang}.} \bibinfo{year}{2023}\natexlab{}.
\newblock \showarticletitle{Deep reinforcement learning in recommender systems:
  A survey and new perspectives}.
\newblock \bibinfo{journal}{\emph{Knowledge-Based Systems}}
  \bibinfo{volume}{264} (\bibinfo{year}{2023}), \bibinfo{pages}{110335}.
\newblock


\bibitem[\protect\citeauthoryear{Chung, Gulcehre, Cho, and Bengio}{Chung
  et~al\mbox{.}}{2014}]%
        {chung2014empirical}
\bibfield{author}{\bibinfo{person}{Junyoung Chung}, \bibinfo{person}{Caglar
  Gulcehre}, \bibinfo{person}{KyungHyun Cho}, {and} \bibinfo{person}{Yoshua
  Bengio}.} \bibinfo{year}{2014}\natexlab{}.
\newblock \showarticletitle{Empirical evaluation of gated recurrent neural
  networks on sequence modeling}.
\newblock \bibinfo{journal}{\emph{arXiv preprint arXiv:1412.3555}}
  (\bibinfo{year}{2014}).
\newblock


\bibitem[\protect\citeauthoryear{Devlin, Chang, Lee, and Toutanova}{Devlin
  et~al\mbox{.}}{2019}]%
        {devlin2019bert}
\bibfield{author}{\bibinfo{person}{Jacob Devlin}, \bibinfo{person}{Ming-Wei
  Chang}, \bibinfo{person}{Kenton Lee}, {and} \bibinfo{person}{Kristina
  Toutanova}.} \bibinfo{year}{2019}\natexlab{}.
\newblock \showarticletitle{BERT: Pre-training of Deep Bidirectional
  Transformers for Language Understanding}. In
  \bibinfo{booktitle}{\emph{Proceedings of the 2019 Conference of the North
  American Chapter of the Association for Computational Linguistics: Human
  Language Technologies, Volume 1 (Long and Short Papers)}}.
  \bibinfo{pages}{4171--4186}.
\newblock


\bibitem[\protect\citeauthoryear{Dubey, Jauhri, Pandey, Kadian, Al-Dahle,
  Letman, Mathur, Schelten, Yang, Fan, et~al\mbox{.}}{Dubey
  et~al\mbox{.}}{2024}]%
        {dubey2024llama}
\bibfield{author}{\bibinfo{person}{Abhimanyu Dubey}, \bibinfo{person}{Abhinav
  Jauhri}, \bibinfo{person}{Abhinav Pandey}, \bibinfo{person}{Abhishek Kadian},
  \bibinfo{person}{Ahmad Al-Dahle}, \bibinfo{person}{Aiesha Letman},
  \bibinfo{person}{Akhil Mathur}, \bibinfo{person}{Alan Schelten},
  \bibinfo{person}{Amy Yang}, \bibinfo{person}{Angela Fan}, {et~al\mbox{.}}}
  \bibinfo{year}{2024}\natexlab{}.
\newblock \showarticletitle{The llama 3 herd of models}.
\newblock \bibinfo{journal}{\emph{arXiv preprint arXiv:2407.21783}}
  (\bibinfo{year}{2024}).
\newblock


\bibitem[\protect\citeauthoryear{Fang, Zhang, Shu, and Guo}{Fang
  et~al\mbox{.}}{2020}]%
        {fang2020deep}
\bibfield{author}{\bibinfo{person}{Hui Fang}, \bibinfo{person}{Danning Zhang},
  \bibinfo{person}{Yiheng Shu}, {and} \bibinfo{person}{Guibing Guo}.}
  \bibinfo{year}{2020}\natexlab{}.
\newblock \showarticletitle{Deep learning for sequential recommendation:
  Algorithms, influential factors, and evaluations}.
\newblock \bibinfo{journal}{\emph{ACM Transactions on Information Systems
  (TOIS)}} \bibinfo{volume}{39}, \bibinfo{number}{1} (\bibinfo{year}{2020}),
  \bibinfo{pages}{1--42}.
\newblock


\bibitem[\protect\citeauthoryear{Feyisetan, Balle, Drake, and Diethe}{Feyisetan
  et~al\mbox{.}}{2020}]%
        {feyisetan2020privacy}
\bibfield{author}{\bibinfo{person}{Oluwaseyi Feyisetan}, \bibinfo{person}{Borja
  Balle}, \bibinfo{person}{Thomas Drake}, {and} \bibinfo{person}{Tom Diethe}.}
  \bibinfo{year}{2020}\natexlab{}.
\newblock \showarticletitle{Privacy-and utility-preserving textual analysis via
  calibrated multivariate perturbations}. In
  \bibinfo{booktitle}{\emph{Proceedings of the 13th international conference on
  web search and data mining}}. \bibinfo{pages}{178--186}.
\newblock


\bibitem[\protect\citeauthoryear{Gan, Wan, and Philip}{Gan
  et~al\mbox{.}}{2023}]%
        {gan2023model}
\bibfield{author}{\bibinfo{person}{Wensheng Gan}, \bibinfo{person}{Shicheng
  Wan}, {and} \bibinfo{person}{S~Yu Philip}.} \bibinfo{year}{2023}\natexlab{}.
\newblock \showarticletitle{Model-as-a-service (MaaS): A survey}. In
  \bibinfo{booktitle}{\emph{2023 IEEE International Conference on Big Data
  (BigData)}}. IEEE, \bibinfo{pages}{4636--4645}.
\newblock


\bibitem[\protect\citeauthoryear{Gao, Fisch, and Chen}{Gao
  et~al\mbox{.}}{2021}]%
        {gao2021making}
\bibfield{author}{\bibinfo{person}{Tianyu Gao}, \bibinfo{person}{Adam Fisch},
  {and} \bibinfo{person}{Danqi Chen}.} \bibinfo{year}{2021}\natexlab{}.
\newblock \showarticletitle{Making Pre-trained Language Models Better Few-shot
  Learners}. In \bibinfo{booktitle}{\emph{Proceedings of the 59th Annual
  Meeting of the Association for Computational Linguistics and the 11th
  International Joint Conference on Natural Language Processing (Volume 1: Long
  Papers)}}. \bibinfo{pages}{3816--3830}.
\newblock


\bibitem[\protect\citeauthoryear{Geng, Liu, Fu, Ge, and Zhang}{Geng
  et~al\mbox{.}}{2022}]%
        {geng2022recommendation}
\bibfield{author}{\bibinfo{person}{Shijie Geng}, \bibinfo{person}{Shuchang
  Liu}, \bibinfo{person}{Zuohui Fu}, \bibinfo{person}{Yingqiang Ge}, {and}
  \bibinfo{person}{Yongfeng Zhang}.} \bibinfo{year}{2022}\natexlab{}.
\newblock \showarticletitle{Recommendation as language processing (rlp): A
  unified pretrain, personalized prompt \& predict paradigm (p5)}. In
  \bibinfo{booktitle}{\emph{Proceedings of the 16th ACM Conference on
  Recommender Systems}}. \bibinfo{pages}{299--315}.
\newblock


\bibitem[\protect\citeauthoryear{Guo, Chen, Zhang, Liu, Dong, and He}{Guo
  et~al\mbox{.}}{2020}]%
        {guo2020leveraging}
\bibfield{author}{\bibinfo{person}{Guibing Guo}, \bibinfo{person}{Bowei Chen},
  \bibinfo{person}{Xiaoyan Zhang}, \bibinfo{person}{Zhirong Liu},
  \bibinfo{person}{Zhenhua Dong}, {and} \bibinfo{person}{Xiuqiang He}.}
  \bibinfo{year}{2020}\natexlab{}.
\newblock \showarticletitle{Leveraging title-abstract attentive semantics for
  paper recommendation}. In \bibinfo{booktitle}{\emph{Proceedings of the AAAI
  conference on artificial intelligence}}, Vol.~\bibinfo{volume}{34}.
  \bibinfo{pages}{67--74}.
\newblock


\bibitem[\protect\citeauthoryear{Guo, Yin, Chen, Zhang, and Zheng}{Guo
  et~al\mbox{.}}{2021}]%
        {guo2021hierarchical}
\bibfield{author}{\bibinfo{person}{Lei Guo}, \bibinfo{person}{Hongzhi Yin},
  \bibinfo{person}{Tong Chen}, \bibinfo{person}{Xiangliang Zhang}, {and}
  \bibinfo{person}{Kai Zheng}.} \bibinfo{year}{2021}\natexlab{}.
\newblock \showarticletitle{Hierarchical hyperedge embedding-based
  representation learning for group recommendation}.
\newblock \bibinfo{journal}{\emph{ACM Transactions on Information Systems
  (TOIS)}} \bibinfo{volume}{40}, \bibinfo{number}{1} (\bibinfo{year}{2021}),
  \bibinfo{pages}{1--27}.
\newblock


\bibitem[\protect\citeauthoryear{Harding, Vanto, Clark, Hannah~Ji, and
  Ainsworth}{Harding et~al\mbox{.}}{2019}]%
        {harding2019understanding}
\bibfield{author}{\bibinfo{person}{Elizabeth~Liz Harding},
  \bibinfo{person}{Jarno~J Vanto}, \bibinfo{person}{Reece Clark},
  \bibinfo{person}{L Hannah~Ji}, {and} \bibinfo{person}{Sara~C Ainsworth}.}
  \bibinfo{year}{2019}\natexlab{}.
\newblock \showarticletitle{Understanding the scope and impact of the
  california consumer privacy act of 2018}.
\newblock \bibinfo{journal}{\emph{Journal of Data Protection \& Privacy}}
  \bibinfo{volume}{2}, \bibinfo{number}{3} (\bibinfo{year}{2019}),
  \bibinfo{pages}{234--253}.
\newblock


\bibitem[\protect\citeauthoryear{Hu, Shen, Wallis, Allen-Zhu, Li, Wang, Wang,
  and Chen}{Hu et~al\mbox{.}}{2021}]%
        {hu2021lora}
\bibfield{author}{\bibinfo{person}{Edward~J Hu}, \bibinfo{person}{Yelong Shen},
  \bibinfo{person}{Phillip Wallis}, \bibinfo{person}{Zeyuan Allen-Zhu},
  \bibinfo{person}{Yuanzhi Li}, \bibinfo{person}{Shean Wang},
  \bibinfo{person}{Lu Wang}, {and} \bibinfo{person}{Weizhu Chen}.}
  \bibinfo{year}{2021}\natexlab{}.
\newblock \showarticletitle{Lora: Low-rank adaptation of large language
  models}.
\newblock \bibinfo{journal}{\emph{arXiv preprint arXiv:2106.09685}}
  (\bibinfo{year}{2021}).
\newblock


\bibitem[\protect\citeauthoryear{Huang, Ma, Wang, Zhou, Yao, Wang, Qi, and
  Chen}{Huang et~al\mbox{.}}{2023}]%
        {huang2023incentive}
\bibfield{author}{\bibinfo{person}{Jiwei Huang}, \bibinfo{person}{Bowen Ma},
  \bibinfo{person}{Ming Wang}, \bibinfo{person}{Xiaokang Zhou},
  \bibinfo{person}{Lina Yao}, \bibinfo{person}{Shoujin Wang},
  \bibinfo{person}{Lianyong Qi}, {and} \bibinfo{person}{Ying Chen}.}
  \bibinfo{year}{2023}\natexlab{}.
\newblock \showarticletitle{Incentive mechanism design of federated learning
  for recommendation systems in mec}.
\newblock \bibinfo{journal}{\emph{IEEE Transactions on Consumer Electronics}}
  (\bibinfo{year}{2023}).
\newblock


\bibitem[\protect\citeauthoryear{Hung, Viet, Tam, Weidlich, Yin, and Zhou}{Hung
  et~al\mbox{.}}{2017}]%
        {hung2017computing}
\bibfield{author}{\bibinfo{person}{Nguyen Quoc~Viet Hung},
  \bibinfo{person}{Huynh~Huu Viet}, \bibinfo{person}{Nguyen~Thanh Tam},
  \bibinfo{person}{Matthias Weidlich}, \bibinfo{person}{Hongzhi Yin}, {and}
  \bibinfo{person}{Xiaofang Zhou}.} \bibinfo{year}{2017}\natexlab{}.
\newblock \showarticletitle{Computing crowd consensus with partial agreement}.
\newblock \bibinfo{journal}{\emph{IEEE Transactions on Knowledge and Data
  Engineering}} \bibinfo{volume}{30}, \bibinfo{number}{1}
  (\bibinfo{year}{2017}), \bibinfo{pages}{1--14}.
\newblock


\bibitem[\protect\citeauthoryear{Jannach and Ludewig}{Jannach and
  Ludewig}{2017}]%
        {jannach2017recurrent}
\bibfield{author}{\bibinfo{person}{Dietmar Jannach} {and}
  \bibinfo{person}{Malte Ludewig}.} \bibinfo{year}{2017}\natexlab{}.
\newblock \showarticletitle{When recurrent neural networks meet the
  neighborhood for session-based recommendation}. In
  \bibinfo{booktitle}{\emph{Proceedings of the eleventh ACM conference on
  recommender systems}}. \bibinfo{pages}{306--310}.
\newblock


\bibitem[\protect\citeauthoryear{Jeckmans, Beye, Erkin, Hartel, Lagendijk, and
  Tang}{Jeckmans et~al\mbox{.}}{2013}]%
        {jeckmans2013privacy}
\bibfield{author}{\bibinfo{person}{Arjan~JP Jeckmans}, \bibinfo{person}{Michael
  Beye}, \bibinfo{person}{Zekeriya Erkin}, \bibinfo{person}{Pieter Hartel},
  \bibinfo{person}{Reginald~L Lagendijk}, {and} \bibinfo{person}{Qiang Tang}.}
  \bibinfo{year}{2013}\natexlab{}.
\newblock \showarticletitle{Privacy in recommender systems}.
\newblock \bibinfo{journal}{\emph{Social media retrieval}}
  (\bibinfo{year}{2013}), \bibinfo{pages}{263--281}.
\newblock


\bibitem[\protect\citeauthoryear{Kang and McAuley}{Kang and McAuley}{2018}]%
        {kang2018self}
\bibfield{author}{\bibinfo{person}{Wang-Cheng Kang} {and}
  \bibinfo{person}{Julian McAuley}.} \bibinfo{year}{2018}\natexlab{}.
\newblock \showarticletitle{Self-attentive sequential recommendation}. In
  \bibinfo{booktitle}{\emph{2018 IEEE international conference on data mining
  (ICDM)}}. IEEE, \bibinfo{pages}{197--206}.
\newblock


\bibitem[\protect\citeauthoryear{Krichene and Rendle}{Krichene and
  Rendle}{2020}]%
        {krichene2020sampled}
\bibfield{author}{\bibinfo{person}{Walid Krichene} {and}
  \bibinfo{person}{Steffen Rendle}.} \bibinfo{year}{2020}\natexlab{}.
\newblock \showarticletitle{On sampled metrics for item recommendation}. In
  \bibinfo{booktitle}{\emph{Proceedings of the 26th ACM SIGKDD international
  conference on knowledge discovery \& data mining}}.
  \bibinfo{pages}{1748--1757}.
\newblock


\bibitem[\protect\citeauthoryear{Lei, Liu, Zhang, Wang, Tang, Li, and Miao}{Lei
  et~al\mbox{.}}{2021}]%
        {lei2021semi}
\bibfield{author}{\bibinfo{person}{Chenyi Lei}, \bibinfo{person}{Yong Liu},
  \bibinfo{person}{Lingzi Zhang}, \bibinfo{person}{Guoxin Wang},
  \bibinfo{person}{Haihong Tang}, \bibinfo{person}{Houqiang Li}, {and}
  \bibinfo{person}{Chunyan Miao}.} \bibinfo{year}{2021}\natexlab{}.
\newblock \showarticletitle{Semi: A sequential multi-modal information transfer
  network for e-commerce micro-video recommendations}. In
  \bibinfo{booktitle}{\emph{Proceedings of the 27th ACM SIGKDD Conference on
  Knowledge Discovery \& Data Mining}}. \bibinfo{pages}{3161--3171}.
\newblock


\bibitem[\protect\citeauthoryear{Li, Wang, Li, Fu, Shen, Shang, and McAuley}{Li
  et~al\mbox{.}}{2023}]%
        {li2023text}
\bibfield{author}{\bibinfo{person}{Jiacheng Li}, \bibinfo{person}{Ming Wang},
  \bibinfo{person}{Jin Li}, \bibinfo{person}{Jinmiao Fu}, \bibinfo{person}{Xin
  Shen}, \bibinfo{person}{Jingbo Shang}, {and} \bibinfo{person}{Julian
  McAuley}.} \bibinfo{year}{2023}\natexlab{}.
\newblock \showarticletitle{Text is all you need: Learning language
  representations for sequential recommendation}. In
  \bibinfo{booktitle}{\emph{Proceedings of the 29th ACM SIGKDD Conference on
  Knowledge Discovery and Data Mining}}. \bibinfo{pages}{1258--1267}.
\newblock


\bibitem[\protect\citeauthoryear{Li, Wen, Wu, Hu, Wang, Li, Liu, and He}{Li
  et~al\mbox{.}}{2021}]%
        {li2021survey}
\bibfield{author}{\bibinfo{person}{Qinbin Li}, \bibinfo{person}{Zeyi Wen},
  \bibinfo{person}{Zhaomin Wu}, \bibinfo{person}{Sixu Hu},
  \bibinfo{person}{Naibo Wang}, \bibinfo{person}{Yuan Li}, \bibinfo{person}{Xu
  Liu}, {and} \bibinfo{person}{Bingsheng He}.} \bibinfo{year}{2021}\natexlab{}.
\newblock \showarticletitle{A survey on federated learning systems: Vision,
  hype and reality for data privacy and protection}.
\newblock \bibinfo{journal}{\emph{IEEE Transactions on Knowledge and Data
  Engineering}} \bibinfo{volume}{35}, \bibinfo{number}{4}
  (\bibinfo{year}{2021}), \bibinfo{pages}{3347--3366}.
\newblock


\bibitem[\protect\citeauthoryear{Liang, Pan, and Ming}{Liang
  et~al\mbox{.}}{2021}]%
        {liang2021fedrec++}
\bibfield{author}{\bibinfo{person}{Feng Liang}, \bibinfo{person}{Weike Pan},
  {and} \bibinfo{person}{Zhong Ming}.} \bibinfo{year}{2021}\natexlab{}.
\newblock \showarticletitle{Fedrec++: Lossless federated recommendation with
  explicit feedback}. In \bibinfo{booktitle}{\emph{Proceedings of the AAAI
  conference on artificial intelligence}}, Vol.~\bibinfo{volume}{35}.
  \bibinfo{pages}{4224--4231}.
\newblock


\bibitem[\protect\citeauthoryear{Lin, Liang, Pan, and Ming}{Lin
  et~al\mbox{.}}{2020}]%
        {lin2020fedrec}
\bibfield{author}{\bibinfo{person}{Guanyu Lin}, \bibinfo{person}{Feng Liang},
  \bibinfo{person}{Weike Pan}, {and} \bibinfo{person}{Zhong Ming}.}
  \bibinfo{year}{2020}\natexlab{}.
\newblock \showarticletitle{Fedrec: Federated recommendation with explicit
  feedback}.
\newblock \bibinfo{journal}{\emph{IEEE Intelligent Systems}}
  \bibinfo{volume}{36}, \bibinfo{number}{5} (\bibinfo{year}{2020}).
\newblock


\bibitem[\protect\citeauthoryear{Lin, Pan, Yang, and Ming}{Lin
  et~al\mbox{.}}{2022}]%
        {lin2022generic}
\bibfield{author}{\bibinfo{person}{Zhaohao Lin}, \bibinfo{person}{Weike Pan},
  \bibinfo{person}{Qiang Yang}, {and} \bibinfo{person}{Zhong Ming}.}
  \bibinfo{year}{2022}\natexlab{}.
\newblock \showarticletitle{A generic federated recommendation framework via
  fake marks and secret sharing}.
\newblock \bibinfo{journal}{\emph{ACM Transactions on Information Systems}}
  \bibinfo{volume}{41}, \bibinfo{number}{2} (\bibinfo{year}{2022}),
  \bibinfo{pages}{1--37}.
\newblock


\bibitem[\protect\citeauthoryear{Liu, Yang, Fan, Peng, and Yu}{Liu
  et~al\mbox{.}}{2022}]%
        {liu2022federated}
\bibfield{author}{\bibinfo{person}{Zhiwei Liu}, \bibinfo{person}{Liangwei
  Yang}, \bibinfo{person}{Ziwei Fan}, \bibinfo{person}{Hao Peng}, {and}
  \bibinfo{person}{Philip~S Yu}.} \bibinfo{year}{2022}\natexlab{}.
\newblock \showarticletitle{Federated social recommendation with graph neural
  network}.
\newblock \bibinfo{journal}{\emph{ACM Transactions on Intelligent Systems and
  Technology (TIST)}} \bibinfo{volume}{13}, \bibinfo{number}{4}
  (\bibinfo{year}{2022}), \bibinfo{pages}{1--24}.
\newblock


\bibitem[\protect\citeauthoryear{Long, Chen, Nguyen, and Yin}{Long
  et~al\mbox{.}}{2023}]%
        {long2023decentralized}
\bibfield{author}{\bibinfo{person}{Jing Long}, \bibinfo{person}{Tong Chen},
  \bibinfo{person}{Quoc Viet~Hung Nguyen}, {and} \bibinfo{person}{Hongzhi
  Yin}.} \bibinfo{year}{2023}\natexlab{}.
\newblock \showarticletitle{Decentralized collaborative learning framework for
  next POI recommendation}.
\newblock \bibinfo{journal}{\emph{ACM Transactions on Information Systems}}
  \bibinfo{volume}{41}, \bibinfo{number}{3} (\bibinfo{year}{2023}),
  \bibinfo{pages}{1--25}.
\newblock


\bibitem[\protect\citeauthoryear{Meng, Huang, Zhang, and Han}{Meng
  et~al\mbox{.}}{2022}]%
        {meng2022generating}
\bibfield{author}{\bibinfo{person}{Yu Meng}, \bibinfo{person}{Jiaxin Huang},
  \bibinfo{person}{Yu Zhang}, {and} \bibinfo{person}{Jiawei Han}.}
  \bibinfo{year}{2022}\natexlab{}.
\newblock \showarticletitle{Generating training data with language models:
  Towards zero-shot language understanding}.
\newblock \bibinfo{journal}{\emph{Advances in Neural Information Processing
  Systems}}  \bibinfo{volume}{35} (\bibinfo{year}{2022}),
  \bibinfo{pages}{462--477}.
\newblock


\bibitem[\protect\citeauthoryear{Min, Ross, Sulem, Veyseh, Nguyen, Sainz,
  Agirre, Heintz, and Roth}{Min et~al\mbox{.}}{2023}]%
        {min2023recent}
\bibfield{author}{\bibinfo{person}{Bonan Min}, \bibinfo{person}{Hayley Ross},
  \bibinfo{person}{Elior Sulem}, \bibinfo{person}{Amir Pouran~Ben Veyseh},
  \bibinfo{person}{Thien~Huu Nguyen}, \bibinfo{person}{Oscar Sainz},
  \bibinfo{person}{Eneko Agirre}, \bibinfo{person}{Ilana Heintz}, {and}
  \bibinfo{person}{Dan Roth}.} \bibinfo{year}{2023}\natexlab{}.
\newblock \showarticletitle{Recent advances in natural language processing via
  large pre-trained language models: A survey}.
\newblock \bibinfo{journal}{\emph{Comput. Surveys}} \bibinfo{volume}{56},
  \bibinfo{number}{2} (\bibinfo{year}{2023}), \bibinfo{pages}{1--40}.
\newblock


\bibitem[\protect\citeauthoryear{Nguyen, Duong, Nguyen, Weidlich, Aberer, Yin,
  and Zhou}{Nguyen et~al\mbox{.}}{2017}]%
        {nguyen2017argument}
\bibfield{author}{\bibinfo{person}{Quoc Viet~Hung Nguyen},
  \bibinfo{person}{Chi~Thang Duong}, \bibinfo{person}{Thanh~Tam Nguyen},
  \bibinfo{person}{Matthias Weidlich}, \bibinfo{person}{Karl Aberer},
  \bibinfo{person}{Hongzhi Yin}, {and} \bibinfo{person}{Xiaofang Zhou}.}
  \bibinfo{year}{2017}\natexlab{}.
\newblock \showarticletitle{Argument discovery via crowdsourcing}.
\newblock \bibinfo{journal}{\emph{The VLDB Journal}}  \bibinfo{volume}{26}
  (\bibinfo{year}{2017}), \bibinfo{pages}{511--535}.
\newblock


\bibitem[\protect\citeauthoryear{Ni, Li, and McAuley}{Ni et~al\mbox{.}}{2019}]%
        {ni2019justifying}
\bibfield{author}{\bibinfo{person}{Jianmo Ni}, \bibinfo{person}{Jiacheng Li},
  {and} \bibinfo{person}{Julian McAuley}.} \bibinfo{year}{2019}\natexlab{}.
\newblock \showarticletitle{Justifying recommendations using distantly-labeled
  reviews and fine-grained aspects}. In \bibinfo{booktitle}{\emph{Proceedings
  of the 2019 conference on empirical methods in natural language processing
  and the 9th international joint conference on natural language processing
  (EMNLP-IJCNLP)}}. \bibinfo{pages}{188--197}.
\newblock


\bibitem[\protect\citeauthoryear{Peng, Yi, Wu, Wu, Zhu, Lyu, Jiao, Xu, Sun, and
  Xie}{Peng et~al\mbox{.}}{2023}]%
        {peng2023you}
\bibfield{author}{\bibinfo{person}{Wenjun Peng}, \bibinfo{person}{Jingwei Yi},
  \bibinfo{person}{Fangzhao Wu}, \bibinfo{person}{Shangxi Wu},
  \bibinfo{person}{Bin Benjamin~Bin Zhu}, \bibinfo{person}{Lingjuan Lyu},
  \bibinfo{person}{Binxing Jiao}, \bibinfo{person}{Tong Xu},
  \bibinfo{person}{Guangzhong Sun}, {and} \bibinfo{person}{Xing Xie}.}
  \bibinfo{year}{2023}\natexlab{}.
\newblock \showarticletitle{Are You Copying My Model? Protecting the Copyright
  of Large Language Models for EaaS via Backdoor Watermark}. In
  \bibinfo{booktitle}{\emph{The 61st Annual Meeting Of The Association For
  Computational Linguistics}}.
\newblock


\bibitem[\protect\citeauthoryear{Qu, Kong, Yang, Zhang, Bendersky, and
  Najork}{Qu et~al\mbox{.}}{2021}]%
        {qu2021natural}
\bibfield{author}{\bibinfo{person}{Chen Qu}, \bibinfo{person}{Weize Kong},
  \bibinfo{person}{Liu Yang}, \bibinfo{person}{Mingyang Zhang},
  \bibinfo{person}{Michael Bendersky}, {and} \bibinfo{person}{Marc Najork}.}
  \bibinfo{year}{2021}\natexlab{}.
\newblock \showarticletitle{Natural language understanding with
  privacy-preserving bert}. In \bibinfo{booktitle}{\emph{Proceedings of the
  30th ACM International Conference on Information \& Knowledge Management}}.
  \bibinfo{pages}{1488--1497}.
\newblock


\bibitem[\protect\citeauthoryear{Radford}{Radford}{2018}]%
        {radford2018improving}
\bibfield{author}{\bibinfo{person}{Alec Radford}.}
  \bibinfo{year}{2018}\natexlab{}.
\newblock \showarticletitle{Improving language understanding by generative
  pre-training}.
\newblock  (\bibinfo{year}{2018}).
\newblock


\bibitem[\protect\citeauthoryear{Ren, Wei, Xia, Su, Cheng, Wang, Yin, and
  Huang}{Ren et~al\mbox{.}}{2024}]%
        {ren2024representation}
\bibfield{author}{\bibinfo{person}{Xubin Ren}, \bibinfo{person}{Wei Wei},
  \bibinfo{person}{Lianghao Xia}, \bibinfo{person}{Lixin Su},
  \bibinfo{person}{Suqi Cheng}, \bibinfo{person}{Junfeng Wang},
  \bibinfo{person}{Dawei Yin}, {and} \bibinfo{person}{Chao Huang}.}
  \bibinfo{year}{2024}\natexlab{}.
\newblock \showarticletitle{Representation learning with large language models
  for recommendation}. In \bibinfo{booktitle}{\emph{Proceedings of the ACM on
  Web Conference 2024}}. \bibinfo{pages}{3464--3475}.
\newblock


\bibitem[\protect\citeauthoryear{Rendle, Freudenthaler, and
  Schmidt-Thieme}{Rendle et~al\mbox{.}}{2010}]%
        {rendle2010factorizing}
\bibfield{author}{\bibinfo{person}{Steffen Rendle}, \bibinfo{person}{Christoph
  Freudenthaler}, {and} \bibinfo{person}{Lars Schmidt-Thieme}.}
  \bibinfo{year}{2010}\natexlab{}.
\newblock \showarticletitle{Factorizing personalized markov chains for
  next-basket recommendation}. In \bibinfo{booktitle}{\emph{Proceedings of the
  19th international conference on World wide web}}. \bibinfo{pages}{811--820}.
\newblock


\bibitem[\protect\citeauthoryear{Sun, Shao, Qian, Huang, and Qiu}{Sun
  et~al\mbox{.}}{2022a}]%
        {sun2022black}
\bibfield{author}{\bibinfo{person}{Tianxiang Sun}, \bibinfo{person}{Yunfan
  Shao}, \bibinfo{person}{Hong Qian}, \bibinfo{person}{Xuanjing Huang}, {and}
  \bibinfo{person}{Xipeng Qiu}.} \bibinfo{year}{2022}\natexlab{a}.
\newblock \showarticletitle{Black-box tuning for language-model-as-a-service}.
  In \bibinfo{booktitle}{\emph{International Conference on Machine Learning}}.
  PMLR, \bibinfo{pages}{20841--20855}.
\newblock


\bibitem[\protect\citeauthoryear{Sun, Xu, Liu, He, Jiang, Wu, and Cui}{Sun
  et~al\mbox{.}}{2022b}]%
        {sun2022survey}
\bibfield{author}{\bibinfo{person}{Zehua Sun}, \bibinfo{person}{Yonghui Xu},
  \bibinfo{person}{Yong Liu}, \bibinfo{person}{Wei He}, \bibinfo{person}{Yali
  Jiang}, \bibinfo{person}{Fangzhao Wu}, {and} \bibinfo{person}{Lizhen Cui}.}
  \bibinfo{year}{2022}\natexlab{b}.
\newblock \showarticletitle{A Survey on Federated Recommendation Systems}.
\newblock \bibinfo{journal}{\emph{arXiv preprint arXiv:2301.00767}}
  (\bibinfo{year}{2022}).
\newblock


\bibitem[\protect\citeauthoryear{Tian, Sun, Poole, Krishnan, Schmid, and
  Isola}{Tian et~al\mbox{.}}{2020}]%
        {tian2020makes}
\bibfield{author}{\bibinfo{person}{Yonglong Tian}, \bibinfo{person}{Chen Sun},
  \bibinfo{person}{Ben Poole}, \bibinfo{person}{Dilip Krishnan},
  \bibinfo{person}{Cordelia Schmid}, {and} \bibinfo{person}{Phillip Isola}.}
  \bibinfo{year}{2020}\natexlab{}.
\newblock \showarticletitle{What makes for good views for contrastive
  learning?}
\newblock \bibinfo{journal}{\emph{Advances in neural information processing
  systems}}  \bibinfo{volume}{33} (\bibinfo{year}{2020}),
  \bibinfo{pages}{6827--6839}.
\newblock


\bibitem[\protect\citeauthoryear{Touvron, Martin, Stone, Albert, Almahairi,
  Babaei, Bashlykov, Batra, Bhargava, Bhosale, et~al\mbox{.}}{Touvron
  et~al\mbox{.}}{2023}]%
        {touvron2023llama}
\bibfield{author}{\bibinfo{person}{Hugo Touvron}, \bibinfo{person}{Louis
  Martin}, \bibinfo{person}{Kevin Stone}, \bibinfo{person}{Peter Albert},
  \bibinfo{person}{Amjad Almahairi}, \bibinfo{person}{Yasmine Babaei},
  \bibinfo{person}{Nikolay Bashlykov}, \bibinfo{person}{Soumya Batra},
  \bibinfo{person}{Prajjwal Bhargava}, \bibinfo{person}{Shruti Bhosale},
  {et~al\mbox{.}}} \bibinfo{year}{2023}\natexlab{}.
\newblock \showarticletitle{Llama 2: Open foundation and fine-tuned chat
  models}.
\newblock \bibinfo{journal}{\emph{arXiv preprint arXiv:2307.09288}}
  (\bibinfo{year}{2023}).
\newblock


\bibitem[\protect\citeauthoryear{Tran and Lauw}{Tran and Lauw}{2022}]%
        {tran2022aligning}
\bibfield{author}{\bibinfo{person}{Nhu-Thuat Tran} {and}
  \bibinfo{person}{Hady~W Lauw}.} \bibinfo{year}{2022}\natexlab{}.
\newblock \showarticletitle{Aligning dual disentangled user representations
  from ratings and textual content}. In \bibinfo{booktitle}{\emph{Proceedings
  of the 28th ACM SIGKDD Conference on Knowledge Discovery and Data Mining}}.
  \bibinfo{pages}{1798--1806}.
\newblock


\bibitem[\protect\citeauthoryear{Vaswani, Shazeer, Parmar, Uszkoreit, Jones,
  Gomez, Kaiser, and Polosukhin}{Vaswani et~al\mbox{.}}{2017}]%
        {vaswani2017attention}
\bibfield{author}{\bibinfo{person}{Ashish Vaswani}, \bibinfo{person}{Noam
  Shazeer}, \bibinfo{person}{Niki Parmar}, \bibinfo{person}{Jakob Uszkoreit},
  \bibinfo{person}{Llion Jones}, \bibinfo{person}{Aidan~N Gomez},
  \bibinfo{person}{{\L}ukasz Kaiser}, {and} \bibinfo{person}{Illia
  Polosukhin}.} \bibinfo{year}{2017}\natexlab{}.
\newblock \showarticletitle{Attention is all you need}.
\newblock \bibinfo{journal}{\emph{Advances in neural information processing
  systems}}  \bibinfo{volume}{30} (\bibinfo{year}{2017}).
\newblock


\bibitem[\protect\citeauthoryear{Voigt and Von~dem Bussche}{Voigt and Von~dem
  Bussche}{2017}]%
        {voigt2017eu}
\bibfield{author}{\bibinfo{person}{Paul Voigt} {and} \bibinfo{person}{Axel
  Von~dem Bussche}.} \bibinfo{year}{2017}\natexlab{}.
\newblock \showarticletitle{The eu general data protection regulation (gdpr)}.
\newblock \bibinfo{journal}{\emph{A Practical Guide, 1st Ed., Cham: Springer
  International Publishing}} \bibinfo{volume}{10}, \bibinfo{number}{3152676}
  (\bibinfo{year}{2017}), \bibinfo{pages}{10--5555}.
\newblock


\bibitem[\protect\citeauthoryear{Wang, Xu, Yang, Shi, Li, Guo, and Liu}{Wang
  et~al\mbox{.}}{2023}]%
        {wang2023knowledge}
\bibfield{author}{\bibinfo{person}{Hao Wang}, \bibinfo{person}{Yao Xu},
  \bibinfo{person}{Cheng Yang}, \bibinfo{person}{Chuan Shi},
  \bibinfo{person}{Xin Li}, \bibinfo{person}{Ning Guo}, {and}
  \bibinfo{person}{Zhiyuan Liu}.} \bibinfo{year}{2023}\natexlab{}.
\newblock \showarticletitle{Knowledge-adaptive contrastive learning for
  recommendation}. In \bibinfo{booktitle}{\emph{Proceedings of the sixteenth
  ACM international conference on web search and data mining}}.
  \bibinfo{pages}{535--543}.
\newblock


\bibitem[\protect\citeauthoryear{Wang, Hu, Wang, Cao, Sheng, and Orgun}{Wang
  et~al\mbox{.}}{2019}]%
        {wang2019sequential}
\bibfield{author}{\bibinfo{person}{Shoujin Wang}, \bibinfo{person}{Liang Hu},
  \bibinfo{person}{Yan Wang}, \bibinfo{person}{Longbing Cao},
  \bibinfo{person}{Quan~Z. Sheng}, {and} \bibinfo{person}{Mehmet Orgun}.}
  \bibinfo{year}{2019}\natexlab{}.
\newblock \showarticletitle{Sequential Recommender Systems: Challenges,
  Progress and Prospects}. In \bibinfo{booktitle}{\emph{Proceedings of the
  Twenty-Eighth International Joint Conference on Artificial Intelligence,
  {IJCAI-19}}}. \bibinfo{publisher}{International Joint Conferences on
  Artificial Intelligence Organization}, \bibinfo{pages}{6332--6338}.
\newblock


\bibitem[\protect\citeauthoryear{Wang, Sun, Xiang, Wu, Ding, Gong, Feng, Shang,
  Zhao, Pang, et~al\mbox{.}}{Wang et~al\mbox{.}}{2021}]%
        {wang2021ernie}
\bibfield{author}{\bibinfo{person}{Shuohuan Wang}, \bibinfo{person}{Yu Sun},
  \bibinfo{person}{Yang Xiang}, \bibinfo{person}{Zhihua Wu},
  \bibinfo{person}{Siyu Ding}, \bibinfo{person}{Weibao Gong},
  \bibinfo{person}{Shikun Feng}, \bibinfo{person}{Junyuan Shang},
  \bibinfo{person}{Yanbin Zhao}, \bibinfo{person}{Chao Pang}, {et~al\mbox{.}}}
  \bibinfo{year}{2021}\natexlab{}.
\newblock \showarticletitle{Ernie 3.0 titan: Exploring larger-scale knowledge
  enhanced pre-training for language understanding and generation}.
\newblock \bibinfo{journal}{\emph{arXiv preprint arXiv:2112.12731}}
  (\bibinfo{year}{2021}).
\newblock


\bibitem[\protect\citeauthoryear{Wang, Zhang, Wang, and Ricci}{Wang
  et~al\mbox{.}}{2022}]%
        {wang2022trustworthy}
\bibfield{author}{\bibinfo{person}{Shoujin Wang}, \bibinfo{person}{Xiuzhen
  Zhang}, \bibinfo{person}{Yan Wang}, {and} \bibinfo{person}{Francesco Ricci}.}
  \bibinfo{year}{2022}\natexlab{}.
\newblock \showarticletitle{Trustworthy recommender systems}.
\newblock \bibinfo{journal}{\emph{ACM Transactions on Intelligent Systems and
  Technology}} (\bibinfo{year}{2022}).
\newblock


\bibitem[\protect\citeauthoryear{Wei, Huang, Xia, and Zhang}{Wei
  et~al\mbox{.}}{2023}]%
        {wei2023multi}
\bibfield{author}{\bibinfo{person}{Wei Wei}, \bibinfo{person}{Chao Huang},
  \bibinfo{person}{Lianghao Xia}, {and} \bibinfo{person}{Chuxu Zhang}.}
  \bibinfo{year}{2023}\natexlab{}.
\newblock \showarticletitle{Multi-modal self-supervised learning for
  recommendation}. In \bibinfo{booktitle}{\emph{Proceedings of the ACM Web
  Conference 2023}}. \bibinfo{pages}{790--800}.
\newblock


\bibitem[\protect\citeauthoryear{Wei, Ren, Tang, Wang, Su, Cheng, Wang, Yin,
  and Huang}{Wei et~al\mbox{.}}{2024}]%
        {wei2024llmrec}
\bibfield{author}{\bibinfo{person}{Wei Wei}, \bibinfo{person}{Xubin Ren},
  \bibinfo{person}{Jiabin Tang}, \bibinfo{person}{Qinyong Wang},
  \bibinfo{person}{Lixin Su}, \bibinfo{person}{Suqi Cheng},
  \bibinfo{person}{Junfeng Wang}, \bibinfo{person}{Dawei Yin}, {and}
  \bibinfo{person}{Chao Huang}.} \bibinfo{year}{2024}\natexlab{}.
\newblock \showarticletitle{Llmrec: Large language models with graph
  augmentation for recommendation}. In \bibinfo{booktitle}{\emph{Proceedings of
  the 17th ACM International Conference on Web Search and Data Mining}}.
  \bibinfo{pages}{806--815}.
\newblock


\bibitem[\protect\citeauthoryear{Wu, Wu, Cao, Huang, and Xie}{Wu
  et~al\mbox{.}}{2021}]%
        {wu2021fedgnn}
\bibfield{author}{\bibinfo{person}{Chuhan Wu}, \bibinfo{person}{Fangzhao Wu},
  \bibinfo{person}{Yang Cao}, \bibinfo{person}{Yongfeng Huang}, {and}
  \bibinfo{person}{Xing Xie}.} \bibinfo{year}{2021}\natexlab{}.
\newblock \showarticletitle{Fedgnn: Federated graph neural network for
  privacy-preserving recommendation}.
\newblock \bibinfo{journal}{\emph{arXiv preprint arXiv:2102.04925}}
  (\bibinfo{year}{2021}).
\newblock


\bibitem[\protect\citeauthoryear{Xie, Sun, Liu, Wu, Gao, Zhang, Ding, and
  Cui}{Xie et~al\mbox{.}}{2022}]%
        {xie2022contrastive}
\bibfield{author}{\bibinfo{person}{Xu Xie}, \bibinfo{person}{Fei Sun},
  \bibinfo{person}{Zhaoyang Liu}, \bibinfo{person}{Shiwen Wu},
  \bibinfo{person}{Jinyang Gao}, \bibinfo{person}{Jiandong Zhang},
  \bibinfo{person}{Bolin Ding}, {and} \bibinfo{person}{Bin Cui}.}
  \bibinfo{year}{2022}\natexlab{}.
\newblock \showarticletitle{Contrastive learning for sequential
  recommendation}. In \bibinfo{booktitle}{\emph{2022 IEEE 38th international
  conference on data engineering (ICDE)}}. IEEE, \bibinfo{pages}{1259--1273}.
\newblock


\bibitem[\protect\citeauthoryear{Yang, Tan, Zheng, Chen, and Yang}{Yang
  et~al\mbox{.}}{2020}]%
        {yang2020federated}
\bibfield{author}{\bibinfo{person}{Liu Yang}, \bibinfo{person}{Ben Tan},
  \bibinfo{person}{Vincent~W Zheng}, \bibinfo{person}{Kai Chen}, {and}
  \bibinfo{person}{Qiang Yang}.} \bibinfo{year}{2020}\natexlab{}.
\newblock \showarticletitle{Federated recommendation systems}.
\newblock \bibinfo{journal}{\emph{Federated Learning: Privacy and Incentive}}
  (\bibinfo{year}{2020}), \bibinfo{pages}{225--239}.
\newblock


\bibitem[\protect\citeauthoryear{Yi, Wu, Wu, Liu, Sun, and Xie}{Yi
  et~al\mbox{.}}{2021}]%
        {yi2021efficient}
\bibfield{author}{\bibinfo{person}{Jingwei Yi}, \bibinfo{person}{Fangzhao Wu},
  \bibinfo{person}{Chuhan Wu}, \bibinfo{person}{Ruixuan Liu},
  \bibinfo{person}{Guangzhong Sun}, {and} \bibinfo{person}{Xing Xie}.}
  \bibinfo{year}{2021}\natexlab{}.
\newblock \showarticletitle{Efficient-FedRec: Efficient Federated Learning
  Framework for Privacy-Preserving News Recommendation}. In
  \bibinfo{booktitle}{\emph{Proceedings of the 2021 Conference on Empirical
  Methods in Natural Language Processing}}. \bibinfo{pages}{2814--2824}.
\newblock


\bibitem[\protect\citeauthoryear{Yin, Qu, Chen, Yuan, Zheng, Long, Xia, Shi,
  and Zhang}{Yin et~al\mbox{.}}{2024}]%
        {yin2024device}
\bibfield{author}{\bibinfo{person}{Hongzhi Yin}, \bibinfo{person}{Liang Qu},
  \bibinfo{person}{Tong Chen}, \bibinfo{person}{Wei Yuan},
  \bibinfo{person}{Ruiqi Zheng}, \bibinfo{person}{Jing Long},
  \bibinfo{person}{Xin Xia}, \bibinfo{person}{Yuhui Shi}, {and}
  \bibinfo{person}{Chengqi Zhang}.} \bibinfo{year}{2024}\natexlab{}.
\newblock \showarticletitle{On-Device Recommender Systems: A Comprehensive
  Survey}.
\newblock \bibinfo{journal}{\emph{arXiv preprint arXiv:2401.11441}}
  (\bibinfo{year}{2024}).
\newblock


\bibitem[\protect\citeauthoryear{Yu, Yin, Xia, Chen, Li, and Huang}{Yu
  et~al\mbox{.}}{2023}]%
        {yu2023self}
\bibfield{author}{\bibinfo{person}{Junliang Yu}, \bibinfo{person}{Hongzhi Yin},
  \bibinfo{person}{Xin Xia}, \bibinfo{person}{Tong Chen},
  \bibinfo{person}{Jundong Li}, {and} \bibinfo{person}{Zi Huang}.}
  \bibinfo{year}{2023}\natexlab{}.
\newblock \showarticletitle{Self-supervised learning for recommender systems: A
  survey}.
\newblock \bibinfo{journal}{\emph{IEEE Transactions on Knowledge and Data
  Engineering}} (\bibinfo{year}{2023}).
\newblock


\bibitem[\protect\citeauthoryear{Yuan, Yang, Nguyen, Cui, He, and Yin}{Yuan
  et~al\mbox{.}}{2023}]%
        {yuan2023interaction}
\bibfield{author}{\bibinfo{person}{Wei Yuan}, \bibinfo{person}{Chaoqun Yang},
  \bibinfo{person}{Quoc Viet~Hung Nguyen}, \bibinfo{person}{Lizhen Cui},
  \bibinfo{person}{Tieke He}, {and} \bibinfo{person}{Hongzhi Yin}.}
  \bibinfo{year}{2023}\natexlab{}.
\newblock \showarticletitle{Interaction-level Membership Inference Attack
  Against Federated Recommender Systems}. In
  \bibinfo{booktitle}{\emph{Proceedings of the ACM Web Conference 2023}}.
  \bibinfo{pages}{1053--1062}.
\newblock


\bibitem[\protect\citeauthoryear{Zhang, Long, Guo, Fang, Song, Liu, Zhou,
  Zhang, Liu, and Yang}{Zhang et~al\mbox{.}}{2024}]%
        {zhang2024federated}
\bibfield{author}{\bibinfo{person}{Chunxu Zhang}, \bibinfo{person}{Guodong
  Long}, \bibinfo{person}{Hongkuan Guo}, \bibinfo{person}{Xiao Fang},
  \bibinfo{person}{Yang Song}, \bibinfo{person}{Zhaojie Liu},
  \bibinfo{person}{Guorui Zhou}, \bibinfo{person}{Zijian Zhang},
  \bibinfo{person}{Yang Liu}, {and} \bibinfo{person}{Bo Yang}.}
  \bibinfo{year}{2024}\natexlab{}.
\newblock \showarticletitle{Federated Adaptation for Foundation Model-based
  Recommendations}.
\newblock \bibinfo{journal}{\emph{arXiv preprint arXiv:2405.04840}}
  (\bibinfo{year}{2024}).
\newblock


\bibitem[\protect\citeauthoryear{Zhang, Xie, Bai, Yu, Li, and Gao}{Zhang
  et~al\mbox{.}}{2021}]%
        {zhang2021survey}
\bibfield{author}{\bibinfo{person}{Chen Zhang}, \bibinfo{person}{Yu Xie},
  \bibinfo{person}{Hang Bai}, \bibinfo{person}{Bin Yu},
  \bibinfo{person}{Weihong Li}, {and} \bibinfo{person}{Yuan Gao}.}
  \bibinfo{year}{2021}\natexlab{}.
\newblock \showarticletitle{A survey on federated learning}.
\newblock \bibinfo{journal}{\emph{Knowledge-Based Systems}}
  \bibinfo{volume}{216} (\bibinfo{year}{2021}), \bibinfo{pages}{106775}.
\newblock


\bibitem[\protect\citeauthoryear{Zhang, Luo, Wu, He, and Li}{Zhang
  et~al\mbox{.}}{2023a}]%
        {zhang2023lightfr}
\bibfield{author}{\bibinfo{person}{Honglei Zhang}, \bibinfo{person}{Fangyuan
  Luo}, \bibinfo{person}{Jun Wu}, \bibinfo{person}{Xiangnan He}, {and}
  \bibinfo{person}{Yidong Li}.} \bibinfo{year}{2023}\natexlab{a}.
\newblock \showarticletitle{LightFR: Lightweight federated recommendation with
  privacy-preserving matrix factorization}.
\newblock \bibinfo{journal}{\emph{ACM Transactions on Information Systems}}
  \bibinfo{volume}{41}, \bibinfo{number}{4} (\bibinfo{year}{2023}),
  \bibinfo{pages}{1--28}.
\newblock


\bibitem[\protect\citeauthoryear{Zhang, Yuan, and Yin}{Zhang
  et~al\mbox{.}}{2023b}]%
        {zhang2023comprehensive}
\bibfield{author}{\bibinfo{person}{Shijie Zhang}, \bibinfo{person}{Wei Yuan},
  {and} \bibinfo{person}{Hongzhi Yin}.} \bibinfo{year}{2023}\natexlab{b}.
\newblock \showarticletitle{Comprehensive privacy analysis on federated
  recommender system against attribute inference attacks}.
\newblock \bibinfo{journal}{\emph{IEEE Transactions on Knowledge and Data
  Engineering}} (\bibinfo{year}{2023}).
\newblock


\bibitem[\protect\citeauthoryear{Zhao, Wang, Xu, Ren, Ng, and Chua}{Zhao
  et~al\mbox{.}}{2024b}]%
        {zhao2024llm}
\bibfield{author}{\bibinfo{person}{Jujia Zhao}, \bibinfo{person}{Wenjie Wang},
  \bibinfo{person}{Chen Xu}, \bibinfo{person}{Zhaochun Ren},
  \bibinfo{person}{See-Kiong Ng}, {and} \bibinfo{person}{Tat-Seng Chua}.}
  \bibinfo{year}{2024}\natexlab{b}.
\newblock \showarticletitle{Llm-based federated recommendation}.
\newblock \bibinfo{journal}{\emph{arXiv preprint arXiv:2402.09959}}
  (\bibinfo{year}{2024}).
\newblock


\bibitem[\protect\citeauthoryear{Zhao, Zhou, Li, Tang, Wang, Hou, Min, Zhang,
  Zhang, Dong, et~al\mbox{.}}{Zhao et~al\mbox{.}}{2023}]%
        {zhao2023survey}
\bibfield{author}{\bibinfo{person}{Wayne~Xin Zhao}, \bibinfo{person}{Kun Zhou},
  \bibinfo{person}{Junyi Li}, \bibinfo{person}{Tianyi Tang},
  \bibinfo{person}{Xiaolei Wang}, \bibinfo{person}{Yupeng Hou},
  \bibinfo{person}{Yingqian Min}, \bibinfo{person}{Beichen Zhang},
  \bibinfo{person}{Junjie Zhang}, \bibinfo{person}{Zican Dong},
  {et~al\mbox{.}}} \bibinfo{year}{2023}\natexlab{}.
\newblock \showarticletitle{A survey of large language models}.
\newblock \bibinfo{journal}{\emph{arXiv preprint arXiv:2303.18223}}
  (\bibinfo{year}{2023}).
\newblock


\bibitem[\protect\citeauthoryear{Zhao, Fan, Li, Liu, Mei, Wang, Wen, Wang,
  Zhao, Tang, et~al\mbox{.}}{Zhao et~al\mbox{.}}{2024a}]%
        {zhao2024recommender}
\bibfield{author}{\bibinfo{person}{Zihuai Zhao}, \bibinfo{person}{Wenqi Fan},
  \bibinfo{person}{Jiatong Li}, \bibinfo{person}{Yunqing Liu},
  \bibinfo{person}{Xiaowei Mei}, \bibinfo{person}{Yiqi Wang},
  \bibinfo{person}{Zhen Wen}, \bibinfo{person}{Fei Wang},
  \bibinfo{person}{Xiangyu Zhao}, \bibinfo{person}{Jiliang Tang},
  {et~al\mbox{.}}} \bibinfo{year}{2024}\natexlab{a}.
\newblock \showarticletitle{Recommender Systems in the Era of Large Language
  Models (LLMs)}.
\newblock \bibinfo{journal}{\emph{IEEE Transactions on Knowledge and Data
  Engineering}} (\bibinfo{year}{2024}).
\newblock


\bibitem[\protect\citeauthoryear{Zheng, Qu, Cui, Shi, and Yin}{Zheng
  et~al\mbox{.}}{2023b}]%
        {zheng2023automl}
\bibfield{author}{\bibinfo{person}{Ruiqi Zheng}, \bibinfo{person}{Liang Qu},
  \bibinfo{person}{Bin Cui}, \bibinfo{person}{Yuhui Shi}, {and}
  \bibinfo{person}{Hongzhi Yin}.} \bibinfo{year}{2023}\natexlab{b}.
\newblock \showarticletitle{Automl for deep recommender systems: A survey}.
\newblock \bibinfo{journal}{\emph{ACM Transactions on Information Systems}}
  \bibinfo{volume}{41}, \bibinfo{number}{4} (\bibinfo{year}{2023}),
  \bibinfo{pages}{1--38}.
\newblock


\bibitem[\protect\citeauthoryear{Zheng, Chen, Wang, Zhao, Yin, and Zhao}{Zheng
  et~al\mbox{.}}{2023a}]%
        {zheng2023multi}
\bibfield{author}{\bibinfo{person}{Shangfei Zheng}, \bibinfo{person}{Wei Chen},
  \bibinfo{person}{Weiqing Wang}, \bibinfo{person}{Pengpeng Zhao},
  \bibinfo{person}{Hongzhi Yin}, {and} \bibinfo{person}{Lei Zhao}.}
  \bibinfo{year}{2023}\natexlab{a}.
\newblock \showarticletitle{Multi-hop knowledge graph reasoning in few-shot
  scenarios}.
\newblock \bibinfo{journal}{\emph{IEEE Transactions on Knowledge and Data
  Engineering}} (\bibinfo{year}{2023}).
\newblock


\bibitem[\protect\citeauthoryear{Zhou, Wang, Zhao, Zhu, Wang, Zhang, Wang, and
  Wen}{Zhou et~al\mbox{.}}{2020}]%
        {zhou2020s3}
\bibfield{author}{\bibinfo{person}{Kun Zhou}, \bibinfo{person}{Hui Wang},
  \bibinfo{person}{Wayne~Xin Zhao}, \bibinfo{person}{Yutao Zhu},
  \bibinfo{person}{Sirui Wang}, \bibinfo{person}{Fuzheng Zhang},
  \bibinfo{person}{Zhongyuan Wang}, {and} \bibinfo{person}{Ji-Rong Wen}.}
  \bibinfo{year}{2020}\natexlab{}.
\newblock \showarticletitle{S3-rec: Self-supervised learning for sequential
  recommendation with mutual information maximization}. In
  \bibinfo{booktitle}{\emph{Proceedings of the 29th ACM international
  conference on information \& knowledge management}}.
  \bibinfo{pages}{1893--1902}.
\newblock


\bibitem[\protect\citeauthoryear{Zhu, Wu, Guo, Hong, and Li}{Zhu
  et~al\mbox{.}}{2024}]%
        {zhu2024collaborative}
\bibfield{author}{\bibinfo{person}{Yaochen Zhu}, \bibinfo{person}{Liang Wu},
  \bibinfo{person}{Qi Guo}, \bibinfo{person}{Liangjie Hong}, {and}
  \bibinfo{person}{Jundong Li}.} \bibinfo{year}{2024}\natexlab{}.
\newblock \showarticletitle{Collaborative large language model for recommender
  systems}. In \bibinfo{booktitle}{\emph{Proceedings of the ACM on Web
  Conference 2024}}. \bibinfo{pages}{3162--3172}.
\newblock


\end{thebibliography}










\end{document}